\documentclass[12pt,preprint,amssymb]{aastex}

\usepackage{graphicx}

\usepackage[figuresright]{rotating}

\def\chandra{{\em Chandra}\/}
\def\asca{{\em ASCA}\/}

\def\rosat{{\em ROSAT}\/}

\def\xmm{{\em XMM-Newton}\/}

\def\arcsec{$^{\prime\prime}$}
\def\arcmin{$^{\prime}$}

\shorttitle{{\it Chandra} Temperature Structure Detections}
\shortauthors{Gu et al.}

\begin{document}
\title{A CHANDRA STUDY OF TEMPERATURE SUBSTRUCTURES IN INTERMEDIATE-REDSHIFT GALAXY CLUSTERS}
\author{Liyi Gu\altaffilmark{1}, Haiguang Xu\altaffilmark{1}, Junhua Gu\altaffilmark{1}, Yu Wang\altaffilmark{1}, Zhongli Zhang\altaffilmark{2}, Jingying Wang\altaffilmark{1}, Zhenzhen Qin\altaffilmark{1}, Haijuan Cui\altaffilmark{1}, Xiang-Ping Wu\altaffilmark{3}}
\altaffiltext{1}{Department of Physics, Shanghai Jiao Tong University,
800 Dongchuan Road,
  Shanghai 200240, PRC}
\altaffiltext{2}{Max-Planck-Institut f\"ur extraterrestrische Physik,
             Giessenbachstra\ss e, 85748 Garching, Germany}
\altaffiltext{3}{National Astronomical Observatories, Chinese Academy of Sciences,
20A Datun Road, Beijing 100012, PRC}

\begin{abstract}

By analyzing the gas temperature maps created from the \chandra\
archive data, we reveal the prevailing existence of temperature
substructures on $\sim 100$ $h_{70}^{-1}$ kpc scales in
the central regions of nine
intermediate-redshift ($z \approx 0.1$) galaxy clusters, which resemble
those found in the Virgo and Coma Clusters. Each substructure contains a
clump of hot plasma whose temperature is about $2 - 3$ keV higher
than the environment, corresponding to an excess thermal energy of
$\sim 10^{58-60}$ erg per clump. Since if there were no significant
non-gravitational heating sources, these substructures would
have perished in $10^{8-9}$ yrs due to thermal
conduction and turbulent flows, whose velocity is found to range from
about 200 to 400 km $\rm s^{-1}$, we conclude
that the substructures cannot be created and sustained by
inhomogeneous radiative cooling. We also eliminate the possibilities
that the temperature substructures are caused by supernova explosions, or by the
non-thermal X-ray emission due to the inverse-Comptonization
of the CMB photons. By calculating the rising time of AGN-induced buoyant bubbles,
we speculate that the intermittent AGN
outbursts ($\geq 10^{60}$ erg per burst) may have played a crucial role in
the forming of the high temperature
substructures. Our results are supported by recent study of McNamara \& Nulsen (2007),
posing a tight observational constraint on future theoretical and numerical studies.

\end{abstract}

\section{INTRODUCTION}

In the frame of hierarchical formation theories, it is believed
that the mass clustering develops due to the gravitational instability
caused by the intrinsic, tiny Gaussian density perturbations that existed in the early
universe (see, e.g., Frenk et al. 1996 for a review). Since the
initial conditions and input physics are assumed to be scale-free,
dark halos are expected to have evolved
self-similarly in time, which can be described by the scaling laws over many
fundamental parameters (Navarro
et al. 1995 and references therein), as have been advocated by the
numerical simulations in which spherical collapse
and adiabatic gas dynamics are assumed (e.g., Navarro et al. 1996; Eke et al. 1998).

On the observational aspect, tight correlations between X-ray luminosity,
gas temperature, gas entropy, X-ray isophotal radius, and virial mass of galaxy clusters have been
discovered successively with {\em Einstein}, {\em ASCA}, and
{\em ROSAT} (e.g., Mushotzky 1984; Mohr \& Evrard 1997; Horner et al. 1999).
These correlations, however, often show notable deviations from theoretical
predictions. For example, in early 1990s, a significant scatter on the measured
luminosity-temperature curves was reported by, e.g., Edge \& Stewart (1991) and Fabian
et al. (1994b), who showed that the cool-core galaxy clusters are more luminous at a given
temperature. The observed mass-temperature relation, on the other hand, shows a lower
($\sim 40\%$)
normalization and a steeper slope with respect to the prediction of adiabatic models (e.g.,
Evrard et al. 1996; Xu et al. 2001). Moreover, an excess in the gas entropy was found
in the central regions ($< 0.05R_{200}$) of many low temperature systems (Ponman et al. 1999, 2003),
which forms a platform of extra energy ($\sim 1-3$ keV per particle; Tozzi \& Norman 2001)
on the entropy-temperature curve.
These deviations and mismatches, which have been confirmed with \chandra\ and \xmm\ observations
recently (e.g., Arnaud et al. 2005; Kotov \& Vikhlinin 2006), clearly indicate that our
understanding of the thermal history of galaxy clusters
is far from complete, and the roles played by non-gravitational heating processes
need to be evaluated properly (e.g., Kaiser 1991; Evrard \& Henry
1991; Kauffmann et al. 1993), which will also help solve the cooling flow problem
(Fabian 1994a; Makishima et al. 2001; Peterson et al. 2003).

As of today, a number of heating sources have been proposed, which include
merger shocks, cosmic rays, supernovae, thermal conduction, turbulent dissipation
and diffusion, and AGN feedbacks (e.g., McNamara \& Nulsen 2007; Markevitch \& Vikhlinin 2007).
However, none of the proposed candidates is all-purpose for
halting the gas cooling and breaking the self-similarities on both galactic
and cluster scales. For example, recent observations (e.g., Croston et al. 2005; Jetha et al. 2007)
showed that, AGN activity,
the most representative heating source of today, tends to deposit most of its energy
into the vast intergalactic space, rather than within the host galaxy ($<0.05 R_{500}$).
Clearly, detailed mappings of gas temperature gradients in the ICM is necessary as a crucial
observational constraint in order to correctly investigate the gas heating history.

Two-dimensional gas temperature variations have been observed in many galaxy clusters with
\chandra\ and {\em XMM-Newton}. By analyzing the
temperature map generated with the high-quality \chandra\ data, Fabian et al. (2000)
identified a pair of cool gas shells that coincide with two radio bubbles, whose radii are
$\sim 6$ and 8 $h_{70}^{-1}$ kpc, respectively, in the innermost region of the Perseus Cluster.
The authors suggested that the cool gas shells were formed as the buoyant gas detached from
the terminals of AGN jets and inflated in pressure equilibrium with the surrounding gas.
Similar phenomenon was also observed in, e.g., Centaurus cluster (Fabian et al. 2005), Abell 2052 (Blanton
et al. 2003), and Hydra A (Nulsen et al. 2002). Also, in
clusters that show clear merger signatures, such as 1E 0657-56 (Govoni et al. 2004), Abell
3667 (Vikhlinin et al. 2001), Abell 665, and Abell 2163 (Markevitch \& Vikhlinin 2001), conspicuous
temperature variations were copiously found, which are presumably caused
by the rapid adiabatic expansion of the shock-heated gas.

So far, however, in most of
the existing works, little has been done to measure the morphologies of the temperature
substructures and to investigate their origins in a quantitative way. One of few exceptional works was
presented by Shibata et al. (2001), who calculated the two point correlation function of the \asca\
hardness ratio map of the Virgo Cluster, and found that there is a characteristic scale of
$\sim 300$ $h_{70}^{-1}$ kpc for the two-dimensional temperature variations. The authors associated this
characteristic scale with the size of the gas blobs arising during the infalling of sub-groups.
By analyzing the two-dimensional temperature distribution with the \xmm\ data, Schuecker et al.
(2004) reported the presence of a turnover at $\sim 100$ $h_{70}^{-1}$ kpc on the pressure
fluctuation spectrum of the Coma Cluster, and attributed it to the existence of Kolmogorov/Oboukhov-type
turbulent flows. As of today, due to the lack of more such quantitative studies, it is not clear
if there does exist a prevalence for a characteristic scale $\sim 100$ $h_{70}^{-1}$ kpc in
galaxy clusters and groups, and how far it can be used to constrain the heating models
in the ICM.

In this work we analyze the {\em Chandra} archive data to search for possible
characteristic scales of temperature variations in a sample of nine intermediate-redshift clusters.
In \S 2 \& 3, we describe the sample selection criteria and data
analysis, respectively. In \S 4, we discuss the physical implications of the results.
In \S 5, we summarize our work. We assume $H_0=70$ km s$^{-1}$
Mpc$^{-1}$, a flat universe for which $\Omega_0=0.3$ and
$\Omega_\Lambda=0.7$, and adopt the solar abundance standards of Grevesse \&
Sauval (1998).

\section{SAMPLE AND DATA PREPARATION}

In order to balance between the angular resolution and detector coverage of the targets,
we construct our sample with nine intermediate-redshift ($z \simeq 0.1$) galaxy clusters,
so that the central 400 $h_{70}^{-1}$ kpc regions of the clusters can be fully, or nearly
fully covered by the S3 CCD of the \chandra\ advanced CCD imaging spectrometer (ACIS)
instrument. All the selected clusters have a $0.7 - 8.0$ keV flux greater than 5.0 $\times
10^{-12}$ erg $\rm s^{-1}$ $\rm cm^{-2}$. We list some basic properties of these
clusters in Table 1, which are arranged in the orders of source
names (col. [1]), names of the cD galaxies (col. [2]), redshifts (col.
[3]), richness classes (col. [4]), right ascension and declination
coordinates (J2000) of the cluster optical centroids (col. [5] \& [6]), and notes (col. [7]).

On the optical and infrared images extracted from the DSS\footnote{http://archive.stsci.edu/dss/} (Fig. 1) and 2MASS\footnote{http://www.ipac.caltech.edu/2mass/}
archives, we do not find any strong evidence for
pronounced signatures caused by major mergers on cluster scales, such as
off-center optical and/or infrared substructures.
A478, A1068, A1650, A2244, and A3112
possess a central galaxy that has been classified as a cD, whose size is
$> 3$ times that of any other member galaxy (Rood \& Sastry 1971; Takizawa et al. 2003).
The remaining four clusters are dominated by a pair or a clump of brightest
member galaxies (Struble \& Rood 1987). In addition, Schombert
et al. (1989) reported the existence of a low-velocity
(the radial velocity difference is $\Delta v_{\rm r} \approx$ 50 km $\rm s^{-1}$)
companion galaxy dwelling deep inside the envelope of the cD galaxy
of A2244.

Strong radio emission associated with AGN activity
has been detected in four clusters. In A478, two
symmetric radio lobes are identified at 1.4 GHz, which extend about
$3^{\prime\prime}$ towards northeast and southwest, respectively
(Sun et al. 2003). In A1068, two compact radio sources are resolved, which
are consistent with the locations of the cD galaxy and SDSS J104043.45+395705.3
(a member galaxy located at about $13^{\prime\prime}$ southwest of the
cluster center), respectively (McNamara et al. 2004).
A2204 and A3112 are reported to possess an extended radio
halo that roughly traces the spatial distribution of the X-ray gas, respectively (Sanders
et al. 2005; Takizawa et al. 2003). In the remaining five clusters, neither radio
lobes nor luminous radio halos has been found in the central region in previous works,
suggesting that they have not experienced strong AGN
outbursts during past few tens of Myr, if the life time of such radio
sources can be estimated by $t_{\rm sync} = \frac{9 m_{\rm e}^3 c^5}{4 e^4 \bar{B}^2 \gamma_{\rm e}}$
and $\bar{B} \sim 10$ $\mu$G (Takahashi \& Yamashita 2003; Donahue et al. 2005).
As a unique case in our sample, A2556 harbors an off-center extended radio
source at about $40^{\prime\prime}$ northeast of the center, which is often speculated
to be the relic of a merger-driven shock (e.g., Govoni et al. 2004).

All the X-ray data analyzed in this work were acquired with the ACIS S3 CCD (Table 2),
with the focal plane temperature set to be --120 $\rm ^{o}C$.
Using the CIAO v4.0 software and CALDB v3.5.0, we removed the bad pixels and columns,
as well as events with \asca\ grades 1, 5, and 7, and then executed the gain, CTI, and
astrometry corrections. In order to identify possible strong background flares,
lightcurves were extracted from regions sufficiently far away, i.e., $\geq 4$\arcmin\ from the X-ray
peaks. Time intervals during which the count rate
exceeds the average quiescent value by 20 percent were excluded. The S1
chip data were used to crosscheck the results when it was available.

\section{X-RAY IMAGING AND SPECTRAL ANALYSIS}

\subsection{X-Ray Images and Surface Brightness Profiles}

Figure 1 shows the intensity contours of the X-ray emission in $0.3-10$ keV
of our sample clusters, which overlay the corresponding optical DSS images.
In all cases, the X-ray morphology appears to be
smooth and symmetric on $> 100$ $h_{70}^{-1}$ kpc scales, showing
no remarkable twists, fronts, or edges that indicate recent major
mergers. On smaller scales, however,
there exist substructures, such as X-ray cavities coinciding with
the AGN lobes (Sun et al. 2003), stripes, filaments,
and arc-like sharp fronts (Wise et al. 2004; Sanders et al. 2005).

For each cluster, we examine the vignetting-corrected surface
brightness profile (SBP) extracted in $0.7-8.0$ keV, after
filtering all the point sources that are detected above the 3$\sigma$
level of the local background by using both the celldetect and
wavdetect tools with their default settings (Fig. 2). By applying the $\chi^2$-test, we find
that, except for the SBP of A2204, which can be well described with
a single-$\beta$ model $S(r) = S(0) (1+(r/r_c)^2)^{-3\beta + 1/2}$,
all the obtained SBPs are consistent with the empirical
two-$\beta$ model $S(r) = S_1(0) (1+(r/r_{c1})^2)^{-3\beta_1 + 1/2} +
S_2(0) (1+(r/r_{c2})^2)^{-3\beta_2 + 1/2} $ (e.g., Ikebe et al. 1996), because
in the central $10 - 20$ $h_{70}^{-1}$ kpc
(A1650, A2556, and A3112) or $20 - 40$ $h_{70}^{-1}$ kpc (A478,
A1068, A1201, A2104, and A2244) there exists an emission excess
beyond the $\beta$ model that can best describe
the SBPs of outer regions. Such an excess is often seen in
clusters that host a central dominating galaxy, which is usually
classified as a cD (Makishima et al. 2001).

\subsection{Azimuthally-averaged Spectral Analysis}
\subsubsection{Background}
In the spectral analysis that follows, we utilized the \chandra\
blank-sky templates for the S3 CCD as the background. The templates
were tailored to match the actual pointings, and the background spectra
were extracted and processed identically to the source spectra (\S 3.2.2).
After rescaling each background spectrum by normalizing its high 
energy end to the corresponding observed spectrum, we further enhance the 
accuracy of the background by adding an unabsorbed APEC model with 
fixed temperature (0.2 keV) and metal abundance (1 $Z_\odot$; Snowden et 
al. 1998) to account for variations in the Galactic X-ray background 
(GXB; Vikhlinin et al. 2005). We note that the GXB component is significant 
only in A2204, where it has a flux of 2.0 $\times 10^{-14}$ erg $\rm s^{-1}$ 
$\rm cm^{-2}$ $\rm arcmin^{-2}$ in $0.3-1.0$ keV and is nearly uniform over 
the S3 chip. This was also reported by Sanders et al. (2009), who indicated
that it is probably due to the excess Galactic foreground.

\subsubsection{Models and Results}
Here we calculate the azimuthally averaged temperature profiles,
which will be used to create the reference temperature maps applied 
in wavelet detection (\S 3.3.2) and calculation of excess energies 
in the temperature substructures (\S 4).
We divided the inner 400 $h_{70}^{-1}$ kpc of each cluster into
concentric annuli, masked all the detected point sources
($\simeq 15$ point sources per cluster), and
extracted the spectrum from each annulus. The corresponding spectral
redistribution matrix files (RMFs) and auxiliary response files (ARFs)
were generated using the CIAO tool mkwarf and mkacisrmf. The spectra
were grouped with $> 20$ counts per bin to allow the use of
$\chi^2$ statistics. We used XSPEC 12.4.0 package to fit the spectra,
by modeling the gas emission
with an absorbed APEC component and employing the PROJCT model
to correct the projection effect. The
absorption column density $N_{\rm{H}}$ was fixed to the Galactic value (Dickey \&
Lockman 1990) in all cases, except for A478, for which $N_{\rm{H}}$ was
set to be about twice the Galactic value (Sanderson et al. 2005; Table 3);
since the absorption excess in A478 exhibits a spatial extension beyond the
cluster's core region, Pointecouteau et al. (2004) argued that it is likely to be associated with a local absorption
substructure in our Galaxy. In order to calculate the error ranges of
the model parameters, we executed the XSPEC command steppar and
carried out multiple iterations at every step, to ensure that the
actual minimum $\chi^2$ is found. The best-fit gas temperature and
abundance profiles are illustrated in Figure 3, in which multiple sets of annuli
are used for each cluster to crosscheck the results. We note that the inclusion of
an additional gas component could only slightly improve the fits for
the central regions of A478, A1650, A2204, A2244, and A3112, which is
insignificant as indicated by the $F$-test.

A significant temperature decline by a
factor of $> 2$ is observed in the central regions of
A478, A1068, A2204, A2556, and A3112, four of which (except A2556)
possess radio halos and/or lobes in the center (\S 2). In the remaining four
clusters without apparent radio emission, the
inward temperature drop is somewhat ambiguous, as the difference between
the core temperature and mean cluster temperature, which refer
to the temperatures measured in the central 100 $h_{70}^{-1}$ kpc
($\sim 0.1r_{500}$) and $100-400$ $h_{70}^{-1}$ kpc regions (Fig. 3), 
respectively, is found within 3$\sigma$ error ranges. If we apply the
classification criteria of Sanderson et al. (2006), the latter four clusters are
non-cooling-core clusters, whose central regions
have not yet fully condensed. These results reflect the
trend for the cooling core clusters to host central radio sources
(Burns 1990). As shown in Figure 3, we also adopt Eqs.$(4)-(6)$ of Vikhlinin et 
al. (2006) to describe the obtained radial temperature profiles 
in a smooth form, which will be used in \S 3.3.2 to calculate the 
reference temperature maps in order to examine the significances of 
detected temperature substructures, and in \S 4 to calculate the 
excess energies in these substructures.


In A478, A1068, A1650, A2204, A2556, and A3112, the
abundance increases inwards significantly (68\% confidence level)
from $\approx 0.2$ $Z_\odot$ in the outermost annuli to $> 0.6$ $Z_\odot$
in the innermost regions ($\lesssim 30$ $h_{70}^{-1}$ kpc).
Emission measure-weighted gas temperature and abundance of each
cluster are also calculated (Table 3), which are
in good agreement with previous results obtained with \chandra\ and
\xmm\ (Sun et al. 2003; Takahashi \& Yamashita 2003; Takizawa et al.
2003; Pointecouteau et al. 2004; Wise et al. 2004; Donahue et al. 2005;
Sanders et al. 2005; Sanderson et al. 2005).

\subsection{Two-Dimensional Analysis}

\subsubsection{Projected Temperature Maps}

Following the methods of, e.g., O'Sullivan et al. (2005), Maughan et
al. (2006), and Kanov et al. (2006), we define a set of knots ${\bf
r_i}$ ($>5000$ knots per cluster) in the central 300 $h_{70}^{-1}$ kpc
region, which are randomly distributed with a separation of $\Delta_{ij}$ $<
5^{\prime\prime}$ between any two adjacent knots $i$ and $j$. To each knot we assign
an adaptively sized circular cell, which is centered on
the knot and contains more than 800 counts in $0.7-8$ keV after the
background is subtracted. The typical cell radius ranges from $\lesssim 5$\arcsec\
at the center to about 10\arcsec\ at r
$\geq 150$ $h_{70}^{-1}$ kpc, which is always larger than the associated $\Delta_{ij}$.
We calculate the projected temperature of each cell $T_{\rm c}({\bf r_i})$,
which is assigned to the corresponding knot,
by fitting the spectrum extracted from the cell with
an absorbed APEC model, using the absorption column given in Table 3;
adding another thermal component can neither improve the fit
significantly, nor make the parameters constrained more tightly. In
the fittings, the source spectrum is grouped to have $> 20$ photons
in each energy bin. In addition to the standard $\chi^2$ statistics,
we also have applied the maximum likelihood statistics, and find
that the best-fit parameters obtained with the two statistics
agree nicely with each other within the $1\sigma$ error ranges. The
calculated $1\sigma$ errors of gas temperature are typically $<10$\% in the
central 100 $h_{70}^{-1}$ kpc, which increase progressively to $\lesssim$ 15\%
(A478, A1650, A2244, and A3112) or $\lesssim$ 20\% (the remaining five
clusters) in $100-300$ $h_{70}^{-1}$ kpc.

For any given position ${\bf r}$, we define a scale $s({\bf r})$, so that there
are at least 800 counts contained in a circular region centered at ${\bf r}$,
whose radius is $s({\bf r})$. We calculate the
projected temperature at ${\bf r}$ by integrating over all the knots ${\bf r_i}$
located in the circular region,
\begin{equation} \label{eq:tmap}
T({\bf r}) = \sum_{\bf r_i} (G_{\bf r_i}(R_{{\bf r},{\bf r_i}})T_{\rm c}({\bf r_i}))/\sum_{\bf r_i} G_{\bf r_i}(R_{{\bf r},{\bf r_i}}), \mbox{ } \rm when \mbox{ } \it R_{{\bf r},{\bf r_i}} < s({\bf r}),
\end{equation}
where $R_{{\bf r},{\bf r_i}}$ is the distance from ${\bf r}$ to ${\bf r_i}$, and $G_{\bf r_i}$ is the Gaussian kernel whose scale
parameter $\sigma$ is fixed at $s({\bf r_i})$. Since $s({\bf r})$ is essentially proportional
to the square root of the local counts, the obtained temperature map $T({\bf r})$ (Fig. 4) is
less affected by the statistical uncertainties caused by surface brightness
fluctuations. In addition, angular resolutions of $\sim 10$\arcsec\ are guaranteed by
the use of compact Gaussian kernel. Maps for the upper
and lower limits of $T({\bf r})$ are calculated in a similar way.

\subsubsection{Temperature Substructures}
\noindent{\bf Wavelet Transform}

\indent A visual inspection of the obtained temperature maps (Fig. 4)
indicates that there exist significant substructures, which typically
have linear sizes of $\sim 100$ $h_{70}^{-1}$ kpc, in all of the
sample clusters. In order to describe these substructures in a quantitative way, we
analyze the statistical properties of the temperature maps by applying the wavelet transform,
a powerful tool in imaging process that identifies and disentangles small
embedded features as a function of scales (e.g., Vikhlinin et al. 1997;
Starck \& Pierre 1998). In an analogue study, Bourdin \& Mazzotta (2008)
conducted the wavelet transform to achieve an optimized spatial binning
of temperature maps. We employ a non-orthogonal version of discrete
wavelet transform, the \`{a} Trous wavelet (Shensa 1992; Starck et
al. 1995; Vikhlinin et al. 1998), which
defines a straightforward decomposition rule. To be specific, at position
${\bf r}=(x,y)$ and scale level $i$ ($i=1,2,...$), the \`{a} Trous wavelet
$w_i(x,y)$ equals the signal difference between two smoothed maps
$c_{i-1}(x,y) - c_i(x,y)$, where $c_{0}(x,y)$ is the obtained temperature
map $T(x,y)$, and $c_{i}(x,y)$ connects with $c_{i-1}(x,y)$ via
$c_{i}(x,y) = \sum_{m,n} h(m,n) c_{i-1}(x+2^{i-1}m,y+2^{i-1}n)$,
in which $h(m,n)$ is the convolution kernel, and $(m,n)$ is its grid.
Here $h(m,n)$ is chosen to be
a compact Gaussian $h(m,n) = e^{-(m^2+n^2)/2/2\pi}$. Each temperature
map is thus decomposed into a set of subimages $w_i(x,y)$ on nine scales
($2^i$ pixels, $i=1,2,..9$), respectively.

Based on the obtained $w_i(x,y)$ maps, we calculate the significant
wavelet coefficient $W_{i}(x,y)$ to detect the temperature
substructures on different scales. For each cluster, we create a set
of random knots ${\bf r_i}$ as used in \S 3.3.1, and assign a
randomized temperature $T_{\rm c}'({\bf r_i})$ to each knot, which
is constrained by the measured temperature profile $T(r)$ and its
error ranges (Fig. 3). By applying Eq.(1) to the $T_{\rm c}'({\bf
r_i})$ map, we obtain a reference temperature map $T_{\rm ref}({\bf
r})$, in which the effects of the radial temperature gradients and
knot tessellation are both involved. A series of such reference
temperature maps are created by repeating this random process, and
each reference map is decomposed into subimages $w_{i,\rm ref}(x,y)$
on nine scales by performing the same \`{a} Trous wavelet transform
as described above. On each scale $i$, we determine the significant
coefficient as $W_{i}(x,y) \equiv w_i(x,y)$ when $w_i(x,y) \geq 3
\sigma_{i,\rm ref}$, and $W_{i}(x,y) \equiv 0$ when $w_i(x,y) < 3
\sigma_{i,\rm ref}$ (e.g., Slezak et al. 1994; Starck et al. 2003),
where $\sigma_{i,\rm ref}$ is the standard deviation of the
simulated $w_{i,\rm ref}(x,y)$ maps so that 3$\sigma_{i,\rm ref}$ is
equivalent to a confidence interval of $1.3 \times 10^{-3}$ for
Gaussian noise. A similar method was used to determine the actual
significances of diffuse sources on \rosat\ images (e.g., Rosati et
al. 1995; Grebenev et al. 1995; Damiani et al. 1997).

Since the \`{a} Trous transform is essentially a non-orthogonal
algorithm, the obtained coefficient $w_i(x,y)$, and thus $W_i(x,y)$, may be
systematically biased from the exact solution (Starck \&
Pierre 1998). This bias can be corrected by using an iterative method (Van
Cittert 1931; Biviano et al. 1996; Starck et al. 2003; Bourdin et al. 2004), the iteration
routine of which is defined as
\begin{equation} \label{eq:vancittert}
W_i^{n}(x,y) = \left \{
\begin {array} {c}
W_i^0(x,y) + W_i^{n-1}(x,y) - PW_i^{n-1}(x,y) \mbox{ } \mbox{ } \mbox{
} \mbox{ } \mbox{ }  W_i^{0}(x,y)\neq0, \\
W_i^{n-1}(x,y)  \mbox{ } \mbox{ } \mbox{ } \mbox{ } \mbox{ } \mbox{ }
\mbox{ } \mbox{ } \mbox{ } \mbox{ }  \mbox{ } \mbox{ } \mbox{ } \mbox{
} \mbox{ } \mbox{ } \mbox{ } \mbox{ } \mbox{ } \mbox{ } \mbox{ }
\mbox{ } \mbox{ } \mbox{ } \mbox{ } \mbox{ } \mbox{ } \mbox{ } \mbox{
} \mbox{ } \mbox{ } \mbox{ } \mbox{ } \mbox{ } \mbox{ } \mbox{ }
\mbox{ } \mbox{ } \mbox{ } \mbox{ } \mbox{ } W_i^{0}(x,y)=0,
\end{array}
\right .
\end{equation}
where $W_i^n(x,y)$ is the corrected detection on scale $i$ after $n$
iteration(s) ($W_i^0(x,y) \equiv W_i(x,y)$ is the initial run),
and $P$ is a non-linear synthesized operator of inverse wavelet
transform, wavelet transform, and filtering operation using
$3\sigma_{i,\rm ref}$ as the threshold. Typically the calculation
converges within five iterations, which yields a corrected $W_i(x,y)$
map as the unbiased reconstruction of the temperature substructures
on scale $i$ (Fig. 4). We define the wavelet spectrum as $\left <\mid W_i(x,y) \mid^2\right >$
(Torrence \& Compo 1998), and plot it in Figure 5 as a function of scale
for each cluster.

As a crosscheck, we also employ the
two-dimensional B-spline function (e.g., Bourdin \& Mazzotta 2008) as the scaling function $\phi(x,y)$
to perform a compact \`{a} Trous transform in the Fourier domain
$(u,v)$, so that the Fourier transform of the wavelet coefficient $w_i(x,y)$
is directly calculated as $\hat{w}_{i}(u,v) =
(1-\hat{h}(2^{i-1}u,2^{i-1}v))\hat{c}_{i-1}(u,v)$, where the filter
$\hat{h}(u,v)=\hat{\phi}(2u,2v)/\hat{\phi}(u,v)$ when $|u|,|v| \leq
1/2$ (in this paper the Fourier transform of any function $f$ is written as
$\hat{f}$). By carrying out the inverse Fourier transform to $\hat{w}_{i}(u,v)$,
${w}_{i}(x,y)$ maps are obtained, and then used to calculate the corrected
$W_i(x,y)$ maps following the procedure described above. The corresponding
wavelet spectra are  illustrated in Figure 5 for comparison, which agree
nicely with those obtained with the Gaussian wavelet. We find that
the most popular scales of the projected temperature substructures are
in the range of about $50 - 200$ $h_{70}^{-1}$ kpc, and about
70\% of these substructures are located at 100 -- 200 $h_{70}^{-1}$ kpc
from the cluster center, exhibiting irregular shapes and spatial
distributions (Fig. 4).

\noindent{\bf Power Spectrum}

\indent As a straightforward approach, we have also studied the power spectra of the two-dimensional gas
temperature distributions. Adopting the method of Schuecker et
al. (2004), we determine the flat fields $\bar{T}({\bf r})$
by applying a low-pass wavelet filter with a scale of $2^8$ pixels
on the original temperature maps,
and calculate the normalized temperature fluctuation distributions as
$\delta T ({\bf r}) = T({\bf r})/\bar{T}({\bf r})-1$.
By carrying out the Fourier transform of $\delta T ({\bf r})$
\begin{equation}
\delta ({\bf k})={1\over S}\int_{S} \delta T({\bf r})e^{i{\bf r}\cdot {\bf k}}d{\bf r},
\end{equation}
where $S$ is the projected area, we obtain the power spectrum $P(k)$,
\begin{equation}
P(k) = \left <\mid{\delta ({\bf k})}\mid^2 \right >.
\end{equation}
Following the method of Schuecker et al. (2004), we carry out 
Monte-Carlo simulations for each cluster to estimate the temperature 
fluctuation at any given knot ${\bf r_i}$ based on the temperature 
error range measured at the same knot (\S 3.3.1). After a randomized 
value $T_{\rm err}({\bf r_i})$ is assigned to each knot, a random 
temperature error map $T_{\rm err}({\bf r})$ is created by applying 
Eq.(1) to the $T_{\rm err}({\bf r_i})$ map, in which the uncertainties 
introduced by data statistics, spectral model fitting, and cell 
tessellation are all involved. Then, by combining $T({\bf r})$ and 
$T_{\rm err}({\bf r})$ maps we create a random temperature map 
$T_{\rm ran}({\bf r})$. The $1\sigma$ error bars shown in Figure 6 are 
calculated from the variances of power spectra obtained from ten such
$T_{\rm ran} ({\bf r})$ maps for each cluster. In general,
the power spectra $P(k)$ show a significant peak in $50 - 200$ $h_{70}^{-1}$
kpc, and drop quickly beyond $200-250$ $h_{70}^{-1}$ kpc. This strongly supports
our conclusion derived with the wavelet analysis (Fig. 5).

\section{DISCUSSION}

We present robust evidence for the prevailing existence of hot gas clumps, whose characteristic scales are $\sim 100$
$h_{70}^{-1}$ kpc, in the central regions of
a sample of nine intermediate-redshift galaxy clusters. These substructures are not caused
by uncertainties in background subtraction, response calibration,
and spectral model fittings, nor are they related to background
flares, the effects of which have been largely corrected and/or fixed. The possibility of 
the temperature substructures arising from the non-thermal emission of low-mass X-ray
binaries (LMXBs) can also be eliminated, since the ICM is only sparsely filled with galaxies. Our results confirm the previous findings in the Coma Cluster
(Schuecker et al. 2004) and the Virgo Cluster (Shibata et al. 2001).

We estimate the excess thermal energy $E_{\rm excess}$ in each
high temperature substructure (Table 4), where the gas temperature is typically
higher than environment by $\Delta T_{\rm X} \approx 2-3$ keV (Fig. 4), by defining the thermal
energy as the product of gas pressure $P$ and volume $V$,
and assuming a spherical or elliptical geometry for the volume of the hot
gas clump. We obtain 
$E_{\rm excess} = \delta P \times V = \Delta T_{\rm X}n_{\rm e}V$
$\sim10^{58-60}$ erg for each clump, where $n_{\rm e}$ is the electron density
determined in the deprojected spectral analysis (\S 3.2.2). To account for
azimuthal variations in the electron density, we include a 10\% systematic 
error (Sanders et al. 2004) in the calculations. As the
projection effect is not corrected for $\Delta T_{\rm X}$, the obtained
excess energy may have been underestimated by a factor of about two. The origin 
of this energy excess is discussed as follows.


\subsection{Inhomogeneous Cooling,  Conduction, and Turbulence}
\indent The detected temperature substructures may have been formed
due to the spatially inhomogeneous cooling in the ICM. Since for a unit volume,
the radiative cooling rate is $R_{\rm rad} = {n_{\rm e}}^2 \Lambda_{\rm rad}(T,Z)$,
where $\Lambda_{\rm rad}$ is the cooling function calculated by
including both continuum and line emissions, the radiative cooling
time of the substructures can be estimated as
\begin{equation} \label{eq:tcool}
 t_{\rm cool} \simeq \frac{n_{\rm e}k_{\rm B}T_{\rm X}}{n_{\rm e}^2
 \Lambda_{\rm rad}}.
\end{equation}
Typically, the calculated cooling time is $\sim 10^{9-10}$ yrs, and
is thus nearly an order of magnitude longer than the characteristic
time for thermal conduction alone to smear out the temperature
substructures (Table 4), which is given by
\begin{equation} \label{eq:ttherm}
t_{\rm cond} \simeq \frac{\delta h}{q} \delta r =  \frac{\frac{5}{2}
  n_{\rm e}k_{\rm B} \delta T_{\rm X}}{f{\rm \kappa_{\rm cond}}\frac{{\rm d}{T_{\rm X}}}{{\rm d}r}}{\rm \delta}r,
\end{equation}
where $\delta h$ is the enthalpy excess per unit volume in the hot
substructure, $q$ is the conductive
flux, ${\rm \delta}r$ is the characteristic scale length of the
substructure, $f$ = 0.2 is the factor to assess the
suppression caused by magnetic fields (e.g., Narayan \& Medvedev 2001),
and $\kappa_{\rm cond}$ is the thermal conductivity of hydrogen plasma,
\begin{equation} \label{eq:conduc}
\kappa_{\rm cond} \simeq 5.0 \times 10^{-7} T_{\rm X}^{\frac{5}{2}}~[{\rm erg} \mbox{ }{\rm cm^{-1}}\mbox{ }{\rm s^{-1}}\mbox{ }{\rm K^{-1}}]
\end{equation}
for Coulomb gas (Spitzer 1962). This indicates that, unless the
conduction process is extremely hindered (i.e., when $f \ll 1$; Maron et
al. 2004; Markevitch et al. 2003), the effect of the radiative cooling
is less important in the formation and evolution of temperature
substructures. Hence the possibility of the substructures being formed by
inhomogeneous cooling can be excluded.

The destruction of a temperature substructure via conduction can be
accelerated by turbulence (Dennis \& Chandran
2005), whose effect can be evaluated by introducing a modified
conductivity (Cho et al. 2003; Voigt \& Fabian 2004)
\begin{equation} \label{eq:turb}
\kappa'_{\rm cond} = \alpha n_{\rm e}k_{\rm B}l\Bigg(\frac {5 k_{\rm B}T}{3 \mu m_{\rm p}}\Bigg)^{1/2},
\end{equation}
where $\alpha$ is the ratio of the turbulence velocity $v_{\rm turb}$
to the local adiabatic sound speed $c_{\rm s}$ ($\alpha = v_{\rm turb}/c_{\rm s}$), and $l$ is the turbulence length
scale. To calculate $\kappa'_{\rm cond}$, the turbulence velocity $v_{\rm turb}$ can be estimated as follows.
As there is evidence that heating continuously compensates a large
fraction of radiative cooling since z$\sim 0.4$ (Bauer et al. 2005; McNamara \& Nulsen 2007), 
we may assume an approximate energy balance between heating and cooling
for the cluster (e.g., Kim \& Narayan 2003; Zakamska \& Narayan
2003; Dennis \& Chandran 2005) as
\begin{equation} \label{eq:balance}
R \simeq H + {\rm \Gamma} + Q,
\end{equation}
where $R$ is the radiative loss, and $H$, $\rm \Gamma$ and $Q$ are the
heating rates of thermal conduction, viscous
dissipation and turbulent diffusion, respectively. Since
\begin{equation} \label{eq:thermalconduction}
H = \nabla \cdot (\kappa_{\rm cond} f \nabla T),
\end{equation}
\begin{equation} \label{eq:turbulentdiss}
{\rm \Gamma} = \frac {c_{\rm diss} \rho v_{\rm turb}^3}{l},
\end{equation}
and
\begin{equation} \label{eq:turbulentdiff}
Q = \nabla \cdot (D \rho T \nabla s),
\end{equation}
where $\rho$ is the gas mass density, $s$ is the specific entropy,
$c_{\rm diss} \approx 0.42$ is a dimensionless constant (Dennis \&
Chandran 2005 and references therein), and $D \approx 10^{29}$ $\rm cm$ $\rm s^{-1}$
is the diffusion coefficient (Rebusco et
al. 2006). By subtracting Eqs.$(10)-(12)$ into Eq.(9), we have
\begin{equation} \label{eq:turbulentvelocity}
v_{\rm turb}=\{[{n_{\rm e}}^2 \Lambda_{\rm rad} -  (\frac {d\Psi}{dr} +
\frac {2\Psi}{r}) - D (\frac{d\Theta}{dr} + \frac
{2\Theta}{r})]\frac{l}{c_{\rm diss} \rho}\}^{\frac{1}{3}},
\end{equation}
if we define $\Psi = f \kappa_{\rm cond} \frac {dT}{dr}$ and $\Theta = \rho T
\frac {ds}{dr}$ (Dennis \& Chandran 2005). Using the observed
deprojected gas density and temperature profiles (Fig. 3), we
calculate $v_{\rm turb}$ and show the results as a function of
turbulence scale $l$ in Figure 7. If we follow
Schuecker et al. (2004) to adopt the lengths indicated by the turnover
points on the wavelet spectra (Fig. 5) as the turbulence scale, we
find that $v_{\rm turb}$ commonly ranges from 200 to 400 $\rm km$ $\rm
s^{-1}$, or $\alpha \sim 0.2 - 0.4$, for our sample. By using Eqs.(6)
and (8), we find that the
inclusion of turbulence reduces the characteristic time $t_{\rm cond}$ to
the modified characteristic time $t'_{\rm cond}$ by a factor
of about 3 (Table 4),
which sets an upper limit ($\sim 10^{8}$ yrs) on the duty cycles of the dominating heating sources.

\subsection{AGN Feedback}
\indent The calculated $E_{\rm excess}$ (Table 4) suggests that the hot clumps
may be dying or remnants of buoyant bubbles, into which the central
AGN has injected energy via shocks and turbulences (e.g., McNamara
\& Nulsen 2007). Such AGN-induced hot gas clumps are
predicted in numerical simulations (e.g., Quilis et al. 2001; Dalla
Vecchia et al. 2004; Vernaleo \& Reynolds 2006) and theoretical calculations (e.g., Pizzolato \&
Soker 2006), whose scales are expected to be about 100 kpc by the
end of the buoyancy process. By examining the temperature maps (Fig. 4),
we note that some hot clumps appear in pairs on both sides
of the central galaxies ($\#$1 and $\#$2 in A1201; $\#$1 and $\#$2 in A2204; $\#$1 and $\#$3 in A3112),
which suggests possible AGN activity taking place at the cluster cores.
However, can an AGN-heated temperature substructure survive
conduction and turbulence destructions? To answer this question, we estimate the characteristic rising
time of a bubble as $t_{\rm buoy} = r_{\rm b} / v_{\rm buoy}$, where $r_{\rm b}$ is
the travel distance measured from the cluster center to the location of a hot clump,
and $v_{\rm buoy}$ is the travel velocity.
In terms of the Keplerian velocity $v_{\rm K}(r_{\rm b})$, a
large bubble is expected to travel through the
ambient gas at
\begin{equation} \label{eq:buoyvelocity}
v_{\rm buoy}(r_{\rm b}) \simeq \sqrt{\frac{8R_{\rm b}}{3Cr_{\rm b}}}v_{\rm K}(r_{\rm b}),
\end{equation}
where $C \simeq 0.75$ is the drag coefficient and $R_{\rm b}$ is
the bubble radius in kpc (Churazov et al. 2001). This yields $t_{\rm buoy}$
smaller or very close to $t'_{\rm cond}$ for $\approx
95$\% of the substructures (Table 4), indicating that such substructures can
survive in most cases. Even for the clumps with $t_{\rm buoy} >
t'_{\rm cond}$, the AGN heating scenario may still stand, if the
distance traveled by the bubble is shorter (e.g., Gardini 2007). It
should be noted that no hot gas clump is found associated with any
known radio source in all sample clusters. This, however, does not
contradict with the AGN heating scenario, because the lack of radio
sources can be explained by the fact that the relativistic electrons
may have lost most of their energy when the bubbles arose, as inferred
by $t_{\rm sync} \lesssim t_{\rm buoy}$ ($t_{\rm sync} \sim
10^{7-8}$ yrs; \S 2).

\subsection{Other Possible Mechanisms}

Numerical and theoretical works show that shocks caused by
supersonic galaxy motion in the ICM can generate observable
temperature substructures via gas compression and friction (e.g.,
El-Zant et al. 2004), and even the low-speed galaxy infalling may also be a
heating source candidate due to the efficient energy conversion via the magnetohydrodynamic turbulence
and magnetic reconnection (Makishima et al. 2001; reviewed by Markevitch \& Vikhlinin
2007). The energy dissipation rate in the central region of
a cluster is as large as $10^{44}$ erg $\rm s^{-1}$ (Fujita et al. 2004),
which assembles $\sim 10^{59}$ erg during the merger process and is thus
sufficient to account for the excess energy in the observed temperature substructures.
Nevertheless, no convincing evidence is given in numerical studies that
the non-filamentary morphologies of the observed temperature substructures can
appear commonly in merger process (e.g., Poole et al. 2006).

We also have estimated the contribution of supernova feedback to the ICM
heating over the substructure lifetime ($t'_{\rm cond}$) using the observed
rate of type Ia supernovae (SNe Ia; Sharon et al. 2007), which implies that
about $0.8-6.1 \times 10^7$ SN Ia
explosions have occurred in the central 400 $h_{70}^{-1}$ kpc region
of each sample cluster. Since the averaged dynamic energy injection into the
environment is $4\times10^{50}$ erg per SN Ia event (Spitzer 1978),
and $\sim 10$ \% of this energy is assumed to have been used to heat the gas
(e.g., Thornton et al. 1998), the total SN Ia heating is found to be $<
3\times 10^{57}$ erg per cluster per $10^8$ yrs. The contribution of type II supernovae
(SNe II) will
not significantly increase the total supernova heating, since even in the IR-luminous
cluster A1068, where the ratio of SN
II rate to SN Ia rate is $\approx 3$ (Wise et al. 2004), the SN II
heating is still lower than $8\times 10^{57}$ erg. Therefore, we conclude that the supernova
heating is insufficient to create the observed high temperature substructures.

High temperature substructures may also occur as a result of the
inverse-Comptonization (IC) of the cosmic microwave background (CMB) photons that are scattered by the relativistic
electrons in the ICM, since in a \chandra\ ACIS spectrum it is difficult to distinguish between
such an IC component and a high temperature thermal component (Petrosian et al. 2008). If in
the ICM $1$\% of the electrons are relativistic ($\gamma \sim 10^{4}$; Eilek 2003),
we will have an IC energy conversion rate of
$b ( \gamma ) = \frac{4}{3} \frac{ \sigma_{\rm T} }{ m_{\rm e} c } \gamma^2 U_{\rm CMB} \simeq 6 \times 10^{44}$
erg $\rm s^{-1}$, where $\sigma_{\rm T}$ is the Thomson cross section
and $U_{\rm CMB}$ is the CMB energy density at the location of the
cluster (Sarazin 1999), which implies that the flux of the IC component is
sufficiently high to bias the temperature measurements by about $1-3$ keV.
However, a tight correlation between the high temperature
substructures and radio synchrotron sources is required in this scenario,
which is actually not observed, possibly due to the lack of relativistic electrons.
Therefore, the IC mechanism for the formation of the temperature substructures
can be eliminated.

\section{SUMMARY}
We reveal the prevailing existence of temperature substructures on
$\sim 100$ $h_{70}^{-1}$ kpc scales in the central regions of nine
intermediate-redshift galaxy clusters. By comparing
the characteristic destruction times of the temperature substructures
with the gas cooling times and the rising times of AGN-induced bubbles,
we conclude that the AGN outbursts may have played a crucial role in
the forming of the observed temperature substructures. Our results agree
with those found earlier in the Virgo and Coma Clusters
(Shibata et al. 2001; Schuecker et al. 2004).

\section*{Acknowledgments}

We thank Kazuo Makishima, Kazuhiro Nakazawa, Peter Schuecker,
and Herv$\acute{\rm e}$ Bourdin for their helpful suggestions and
comments. This work was supported by the National Science Foundation of 
China (Grant No. 10673008 and 10878001), the Ministry 
of Science and Technology of China (Grant No. 2009CB824900/2009CB24904),
and the Ministry of Education of China (the NCET Program).

\clearpage

\begin{deluxetable}{clccccrcc}

\tabletypesize{\scriptsize} \tablewidth{0pt} \tablecaption{Galaxy Clusters in Our Sample}
\tablehead{ \colhead{Name}           & \colhead{cD Galaxy} &
\colhead{Redshift}       & \colhead{R\tablenotemark{a}}       &
\colhead{RA\tablenotemark{b}}             & \colhead{DEC\tablenotemark{b}}     &
\colhead{Notes\tablenotemark{c}} \\
\colhead{} & \colhead{} & \colhead{} & \colhead{} & \colhead{(h d s;
J2000)} & \colhead{(d m s; J2000)} & \colhead{}
 }
\startdata

A0478 & 2MASX J04132526+1027551 & 0.0881& 2& 04 13 20.7 & +10 28 35  & 1,3 \\

A1068 & 2MASX J10404446+3957117 & 0.1375& 1& 10 40 47.1 & +39 57 19  & 2,3 \\

A1201 & $-$ & 0.1688& 2& 11 13 01.1 & +13 25 40 & \\

A1650 & 2MASX J12584149--0145410 & 0.0845& 2& 12 58 46.2 & $-$01 45 11  & \\

A2104 & $-$ & 0.1554& 2& 15 40 06.8 & $-$03 17 39 & \\

A2204 & $-$ & 0.1523& 3& 16 32 45.7 & +05 34 43 & 2,3  \\

A2244 & FIRST J170242.5+340337 & 0.0968& 2& 17 02 44.0 & +34 02 48  & \\

A2556 & $-$ & 0.0871& 1& 23 13 01.6 & $-$21 37 59  & 4 \\

A3112 & ESO 248- G 006 & 0.0750& 2& 03 17 52.4 & $-$44 14 35 & 3 \\
\enddata

\tablenotetext{a}{Abell richness class (Abell et al. 1989).}
\tablenotetext{b}{Positions of the cluster optical centroids.}
\tablenotetext{c}{X-ray and radio substructures in the clusters, including 1 $-$ X-ray cavities, 2 $-$ cold fronts, 3 $-$ central radio sources, and 4 $-$ off-center radio source.}
\end{deluxetable}

\clearpage

\begin{deluxetable}{llllcc}
\tabletypesize{\scriptsize} \tablecaption{Observation Log
\label{tbl:ObsLog}} \tablewidth{0pt} \tablecolumns{10} \tablehead{
\colhead{Target} & \colhead{ObsID} & \colhead{ACIS CCD} &
\colhead{Mode} & \colhead{Date} &
\colhead{Raw/Clean Exposure}\\
\colhead{} & \colhead{} & \colhead{} & \colhead{} & \colhead{dd mm
yyyy} & \colhead{(ks)} } \startdata
A0478 & 1669 & 235678 & FAINT  & 27/01/2001 & 42.9/41.9 \\
A1068 & 1652 & 236789 & FAINT  & 04/02/2001 & 27.2/26.0 \\
A1201 & 4216 & 35678  & VFAINT & 01/11/2003 & 40.2/25.4 \\
A1650 & 4178 & 35678  & VFAINT & 03/08/2003 & 27.6/26.0 \\
A2104 & 895  & 235678 & FAINT  & 25/05/2000 & 49.8/48.1 \\
A2204 & 499  & 235678 & FAINT  & 29/07/2000 & 10.2/9.3  \\
A2244 & 4179 & 35678  & VFAINT & 10/10/2003 & 57.7/56.0 \\
A2556 & 2226 & 35678  & VFAINT & 05/10/2001 & 20.2/19.8 \\
A3112 & 2516 & 35678  & VFAINT & 15/09/2001 & 17.2/15.1 \\
\enddata
\end{deluxetable}

\clearpage

\begin{deluxetable}{cccccc}
\centering \tabletypesize{\scriptsize} \tablewidth{0pt}
\tablecaption{Global Properties of the Halo Gas\tablenotemark{a}}
\tablehead{ \colhead{Name}           & \colhead{Aperture}      &
\colhead{$kT$}         & \colhead{Abundance}     &
\colhead{$N_{\rm{H}}$\tablenotemark{b}} \\
\colhead{} & \colhead{($h_{70}^{-1}$ kpc)} & \colhead{(keV)} &
\colhead{($Z_\odot$)} & \colhead{$(10^{20}$ cm$^{-2})$}
   }

\startdata

A0478 & 405.4 & $6.93^{+0.20}_{-0.26}$& $0.47^{+0.03}_{-0.03}$& $26.75^{+0.72}_{-0.51}$ \\

A1068 & 418.7& $3.63^{+0.10}_{-0.10}$& $0.56^{+0.04}_{-0.04}$& 1.40\\

A1201 & 425.5& $4.81^{+0.35}_{-0.44}$& $0.32^{+0.11}_{-0.09}$& 1.61\\

A1650 & 390.5& $5.96^{+0.24}_{-0.25}$& $0.48^{+0.07}_{-0.08}$& 1.56\\

A2104 & 397.6& $9.60^{+0.77}_{-0.76}$& $0.47^{+0.07}_{-0.08}$& 8.69\\

A2204 & 391.0& $7.24^{+0.43}_{-0.53}$& $0.51^{+0.06}_{-0.07}$& 5.67\\

A2244 & 396.9& $5.25^{+0.10}_{-0.08}$& $0.29^{+0.03}_{-0.03}$& 2.14\\

A2556 & 401.3& $3.17^{+0.09}_{-0.10}$& $0.43^{+0.05}_{-0.04}$& 2.05\\

A3112 & 385.5& $4.03^{+0.13}_{-0.12}$& $0.79^{+0.08}_{-0.07}$& 2.61\\

\enddata
\tablenotetext{a}{An absorbed APEC model is used to estimate the emission measure-weighted
gas temperatures $kT$ and abundances. The error bars are given at 90\% confidence level.}
\tablenotetext{b}{Column densities are fixed to the Galactic values (Dickey \& Lockman 1990), except for
A478, which shows an absorption larger than the Galactic value (15.1 $\times 10^{20}$ $\rm cm^{-2}$).}

\end{deluxetable}

\clearpage

\clearpage

\clearpage
\begin{deluxetable}{cccccccccc}
\tabletypesize{\scriptsize} \tablewidth{0pt}
\tablecaption{Characteristic Timescales for the Detected Temperature Substructures}
\tablehead{ \colhead{No.\tablenotemark{a}}           &
\colhead{$r$ \tablenotemark{b}}      & \colhead{$\rm n_e$\tablenotemark{c}}
& \colhead{$\rm E_{excess}$ \tablenotemark{d}}     & \colhead{$\rm
t_{cool}$ \tablenotemark{e}} & \colhead{$\rm t_{cond}$
\tablenotemark{f}} & \colhead{$\rm t^{'}_{cond}$ \tablenotemark{g}} &
\colhead{$\rm t_{buoy}$ \tablenotemark{h}}
\\
\colhead{} & \colhead{($h_{70}^{-1}$ kpc)} & \colhead{($\rm
cm^{-3}$)} & \colhead{($10^{58}$ erg)} & \colhead{($10^8$ yrs)} &
\colhead{($10^8$ yrs)} & \colhead{($10^8$ yrs)} & \colhead{($10^8$
yrs)} &

   }
\startdata
A0478[1] & 95& 0.016& ${1.9 \pm 0.8}$& ${16.4 \pm 5.2}$ & ${7.7 \pm 2.0}$ & ${1.9 \pm 0.4}$ & 1.6 \\

\hspace{7 mm} [2] & 141& 0.010& ${7.4 \pm 2.2}$& ${23.8 \pm 6.1}$ & ${3.2 \pm 0.8}$ & ${1.7 \pm 0.6}$ & 2.0 \\

\hspace{7 mm} [3] & 155& 0.009& ${7.2 \pm 3.7}$& ${30.0 \pm 7.4}$ & ${2.4 \pm 0.4}$ & ${1.7 \pm 0.6}$ & 2.2 \\

\hspace{7 mm} [4] & 157& 0.009& ${4.3 \pm 2.5}$& ${33.6 \pm 8.1}$ & ${4.7 \pm 0.8}$ & ${2.3 \pm 0.5}$ & 2.3 \\

\hline

A1068[1] & 184& 0.004& ${11.0 \pm 4.8}$& ${54.6 \pm 13.8}$ & ${16.7 \pm 2.2}$ & ${2.2 \pm 0.8}$ & 2.7 \\

\hspace{7 mm} [2] & 237& 0.002& ${14.6 \pm 8.7}$& ${106.2 \pm 19.8}$ & ${8.7 \pm 2.2}$ & ${5.4 \pm 1.5}$ & 3.2 \\

\hspace{7 mm} [3] & 244& 0.002& ${32.2 \pm 15.8}$& ${135.9 \pm 30.4}$ & ${10.3 \pm 2.9}$ & ${5.3 \pm 1.6}$ & 3.3 \\

\hline

A1201[1] & 123& 0.005& ${17.4 \pm 10.8}$& ${69.1 \pm 16.7}$ & ${5.8 \pm 1.2}$ & ${4.2 \pm 0.9}$ & 2.0 \\

\hspace{7 mm} [2] & 171& 0.003& ${36.0 \pm 20.7}$& ${87.6 \pm 17.8}$ & ${11.1 \pm 2.1}$ & ${5.9 \pm 1.4}$ & 2.7 \\

\hspace{7 mm} [3] & 185& 0.003& ${29.3 \pm 16.7}$& ${118.4 \pm 30.5}$ & ${9.2 \pm 1.7}$ & ${4.7 \pm 1.0}$ & 2.9 \\

\hline

A1650[1] &  72& 0.013& ${4.6 \pm 2.0}$& ${26.1 \pm 12.0}$ & ${3.2 \pm 1.2}$ & ${1.5 \pm 0.8}$ & 2.4 \\

\hspace{7 mm} [2] & 108& 0.007& ${12.2 \pm 5.2}$& ${42.3 \pm 12.9}$ & ${6.3 \pm 2.3}$ & ${2.2 \pm 0.5}$ & 2.9 \\

\hspace{7 mm} [3] & 113& 0.007& ${4.6 \pm 1.5}$& ${38.7 \pm 12.2}$ & ${4.7 \pm 1.4}$ & ${2.3 \pm 0.5}$ & 3.0 \\

\hspace{7 mm} [4] & 171& 0.005& ${12.6 \pm 4.5}$& ${61.5 \pm 13.6}$ & ${2.8 \pm 1.1}$ & ${2.5 \pm 1.1}$ & 3.6\\

\hline

A2104[1] &  132& 0.007& ${7.0 \pm 5.6}$ & ${53.7 \pm 16.8}$ & ${3.4 \pm 0.7}$ & ${2.7 \pm 0.5}$& 2.8 \\

\hspace{7 mm} [2] & 253& 0.003& ${47.8 \pm 26.5}$& ${87.7 \pm 17.8}$ & ${1.8 \pm 0.5}$ & ${0.4 \pm 0.3}$ & 4.0 \\

\hspace{7 mm} [3] & 261& 0.003& ${41.5 \pm 23.9}$& ${122.8 \pm 31.2}$ & ${1.7 \pm 0.5}$ & ${0.2 \pm 0.3}$ & 4.1 \\

\hline

A2204[1] & 94& 0.017& ${65.8 \pm 26.1}$& ${22.8 \pm 6.5}$ & ${1.5 \pm 0.5}$ & ${1.4 \pm 0.5}$& 1.4 \\

\hspace{7 mm} [2] & 117& 0.013& ${54.0 \pm 26.8}$& ${22.8 \pm 6.8}$ & ${3.5 \pm 0.6}$ & ${1.9 \pm 0.5}$ & 1.6\\

\hspace{7 mm} [3] & 172& 0.009& ${145.7 \pm 72.3}$& ${47.2 \pm 11.7}$ & ${6.0 \pm 1.6}$ & ${2.9 \pm 1.0}$ & 2.0\\

\hspace{7 mm} [4] & 220& 0.006& ${104.4 \pm 42.1}$& ${61.8 \pm 13.9}$ & ${2.8 \pm 0.9}$ & ${2.1 \pm 0.8}$ & 2.4\\
\hline

A2244[1] &  74& 0.014& ${3.6 \pm 1.0}$& ${22.7 \pm 6.1}$ & ${15.0 \pm 0.9}$ & ${2.5 \pm 0.5}$& 1.3 \\

\hspace{7 mm} [2] & 100& 0.010& ${8.5 \pm 3.2}$& ${29.8 \pm 7.8}$ & ${8.9 \pm 0.7}$ & ${2.5 \pm 0.5}$ & 1.7 \\

\hspace{7 mm} [3] & 155& 0.006& ${4.9 \pm 2.1}$& ${42.0 \pm 13.0}$ & ${6.4 \pm 0.8}$ & ${2.5 \pm 0.6}$ & 2.3 \\

\hline

A2556[1] & 108& 0.005& ${13.6 \pm 3.8}$& ${62.3 \pm 12.8}$ & ${13.9 \pm 2.1}$ & ${3.7 \pm 0.8}$& 1.8 \\

\hspace{7 mm} [2] & 141& 0.004& ${13.0 \pm 2.2}$& ${70.8 \pm 12.0}$ & ${10.7 \pm 1.5}$ & ${3.2 \pm 0.7}$ & 2.3 \\

\hspace{7 mm} [3] & 144& 0.004& ${7.0 \pm 3.6}$& ${65.3 \pm 15.4}$ & ${16.0 \pm 2.6}$ & ${3.6 \pm 0.9}$ & 2.3 \\

\hline

A3112[1] &  73& 0.014& ${2.9 \pm 1.3}$& ${20.2 \pm 6.4}$ & ${9.5 \pm 1.1}$ & ${1.9 \pm 0.6}$& 1.2 \\

\hspace{7 mm} [2] & 121& 0.006& ${5.2 \pm 2.2}$& ${45.1 \pm 9.1}$ & ${11.4 \pm 1.5}$ & ${2.9 \pm 0.7}$ & 1.8 \\

\hspace{7 mm} [3] & 125& 0.006& ${6.8 \pm 3.6}$& ${41.9 \pm 9.5}$ & ${7.2 \pm 1.2}$ & ${2.2 \pm 0.6}$ & 1.9 \\

\hspace{7 mm} [4] & 141& 0.006& ${3.8 \pm 2.2}$& ${39.6 \pm 9.6}$ & ${2.8 \pm 0.8}$ & ${2.1 \pm 0.6}$ & 2.0 \\

\enddata
\tablenotetext{a}{Detected temperature substructures. See Figure 4.}
\tablenotetext{b}{Distances between the geometric centers of the
substructures and the cluster centers.} \tablenotetext{c}{Averaged electron densities of
the substructures.} \tablenotetext{d}{Excess
thermal energies contained in the substructures.} \tablenotetext{e}{Radiative cooling
times of the substructures.}
\tablenotetext{f}{Destruction times of the substructures due to
thermal conduction (Eq.(6)).} \tablenotetext{g}{Modified destruction times of the substructures
due to thermal conduction, viscous dissipation, and turbulent diffusion (Eq.(8)).}
\tablenotetext{h}{Rising times of the temperature substructures if they are AGN-induced bubbles.}
\end{deluxetable}

\clearpage

\clearpage

\begin{figure}
\begin{center}
\includegraphics[scale=.8]{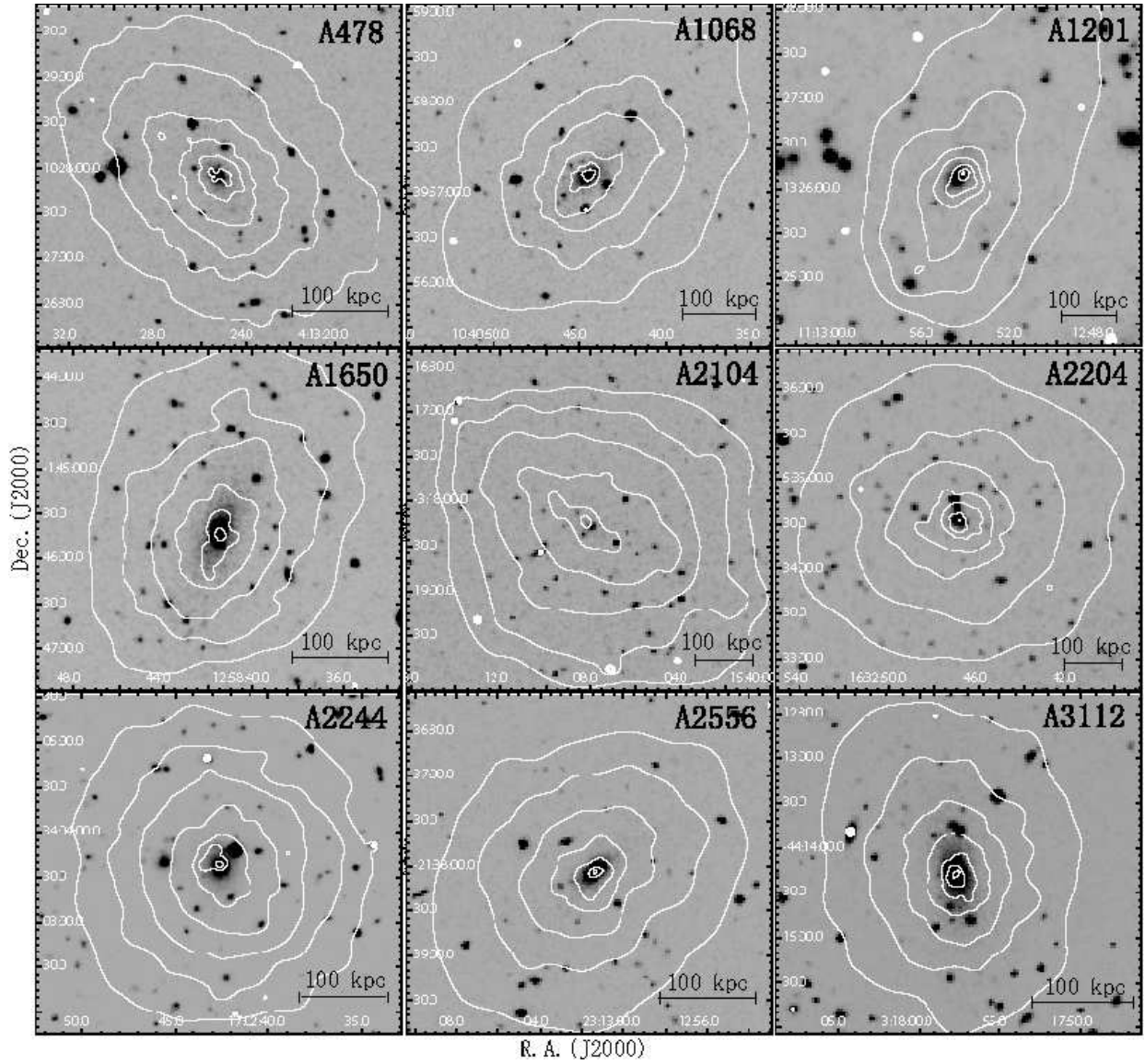}
\caption{DSS optical images of the sample clusters, on which the $0.3-10.0$ keV
X-ray intensity contours are plotted. To calculate the X-ray contours we subtract the background,
correct the exposure, and smooth the image with a minimum signal-to-noise (S/N) of 3
and a maximum S/N of 5 per smoothing beam.}
\end{center}
\end{figure}

\begin{figure}
\epsscale{1.0}
\begin{center}
\includegraphics[angle=0,scale=.70]{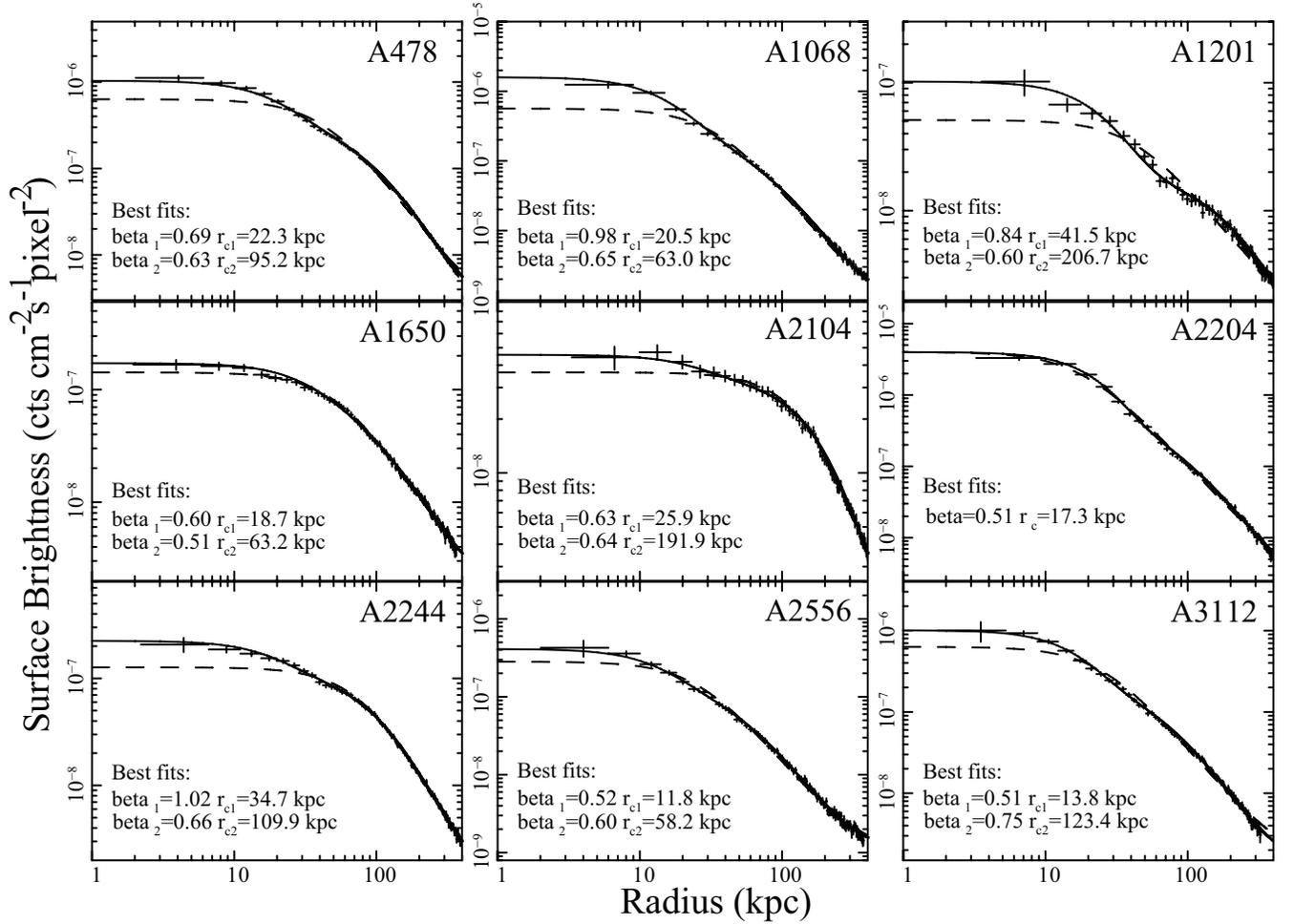}
\caption{ Radial surface brightness profiles extracted in $0.7-8.0$ keV, which are
corrected for both background and exposure. The
best-fit two--$\beta$ and $\beta$ models are shown with solid
and dash lines, respectively. For A2204 only the $\beta$ model is plotted, which is sufficient
to describe its SBP.}
\end{center}
\end{figure}

\begin{figure}
\begin{center}
\includegraphics[angle=0,scale=.55]{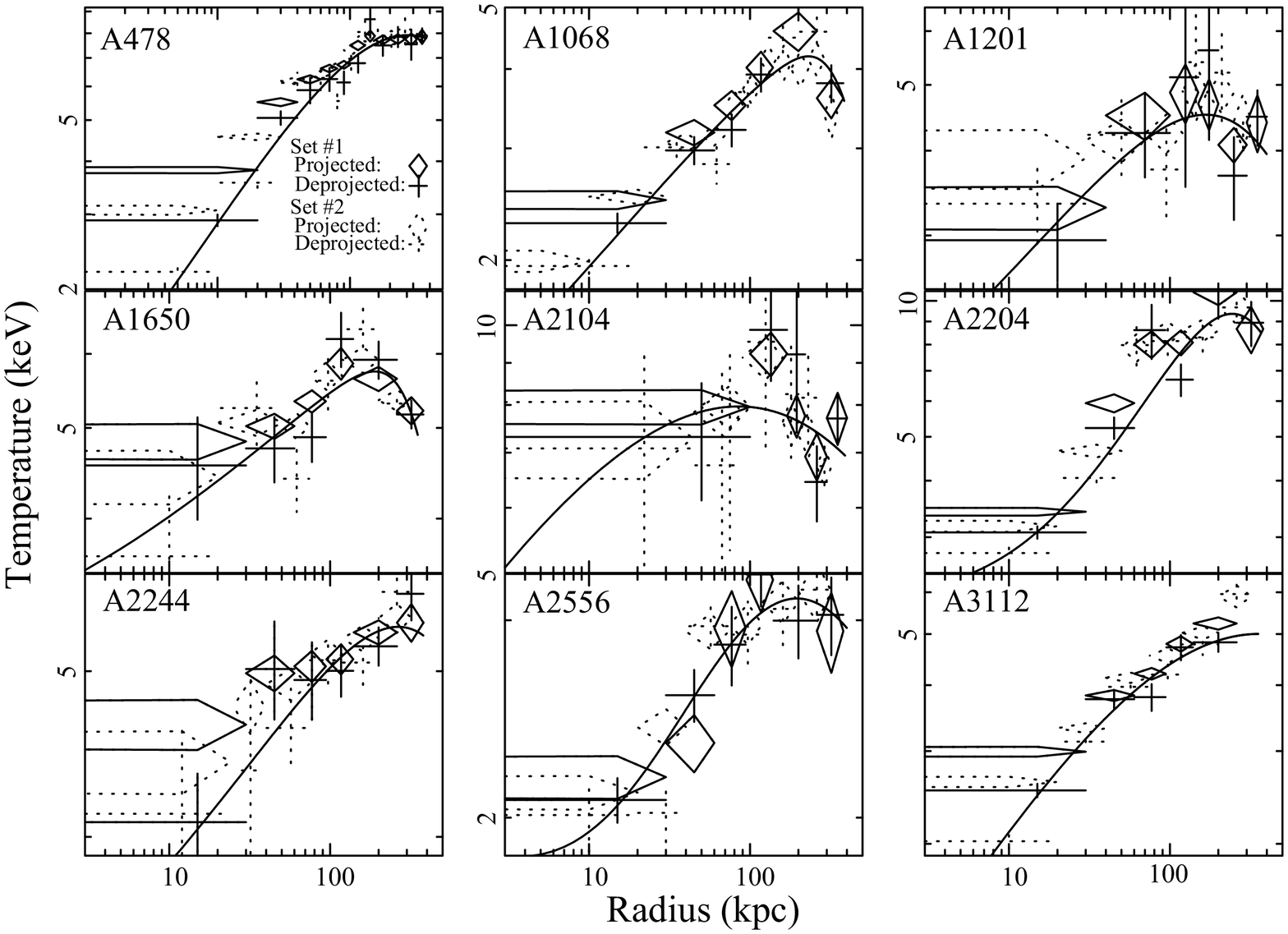}
\includegraphics[angle=0,scale=.55]{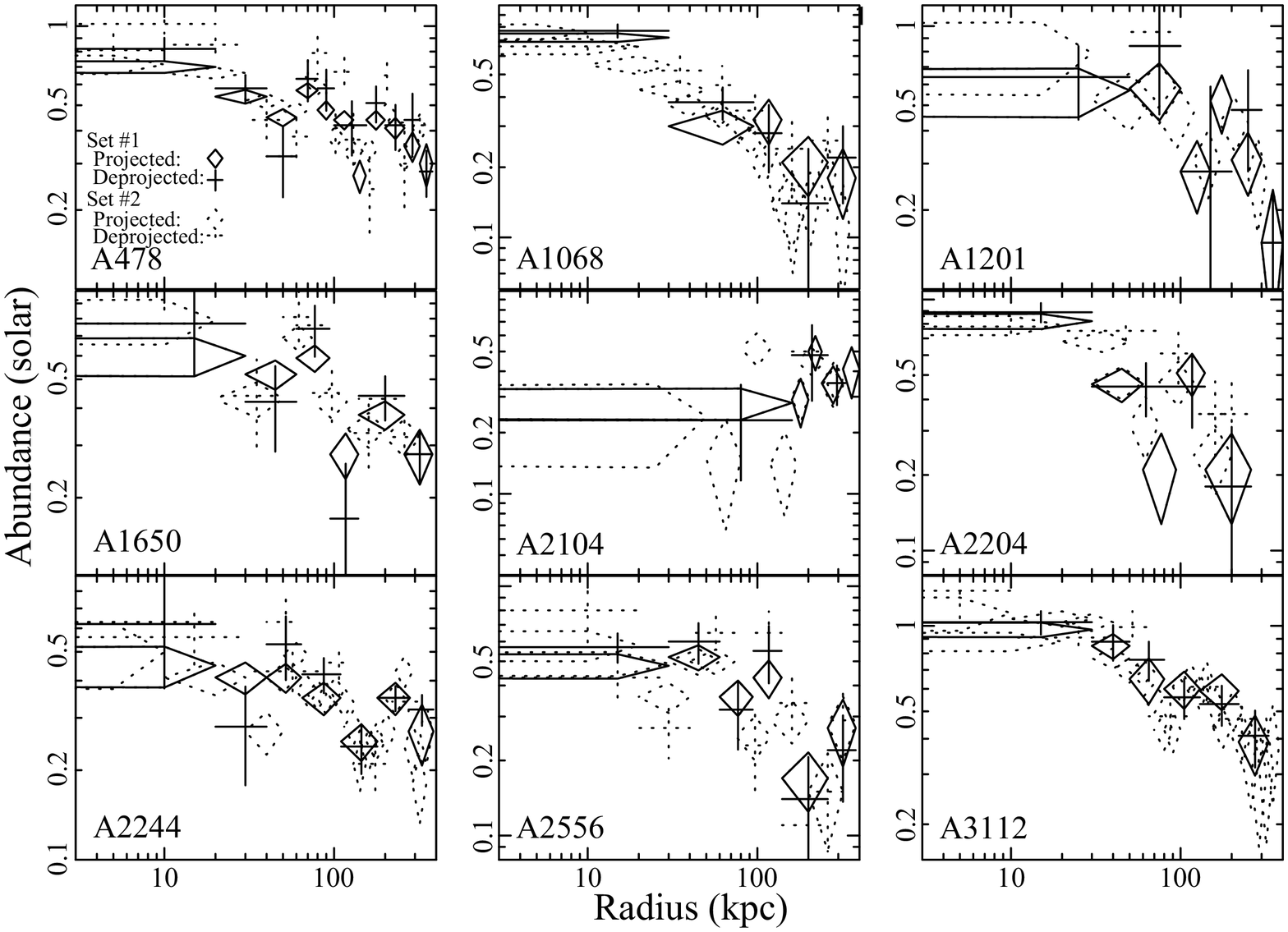}
\caption{Projected and deprojected temperature ({\it upper
panel}) and abundance profiles ({\it lower panel}) for two sets of annuli.
Error bars are given at 90\% and 68\% confidence levels
for temperature and abundance profiles, respectively.}

\end{center}
\end{figure}

\begin{figure}
\begin{center}
\includegraphics[angle=0,scale=1.5]{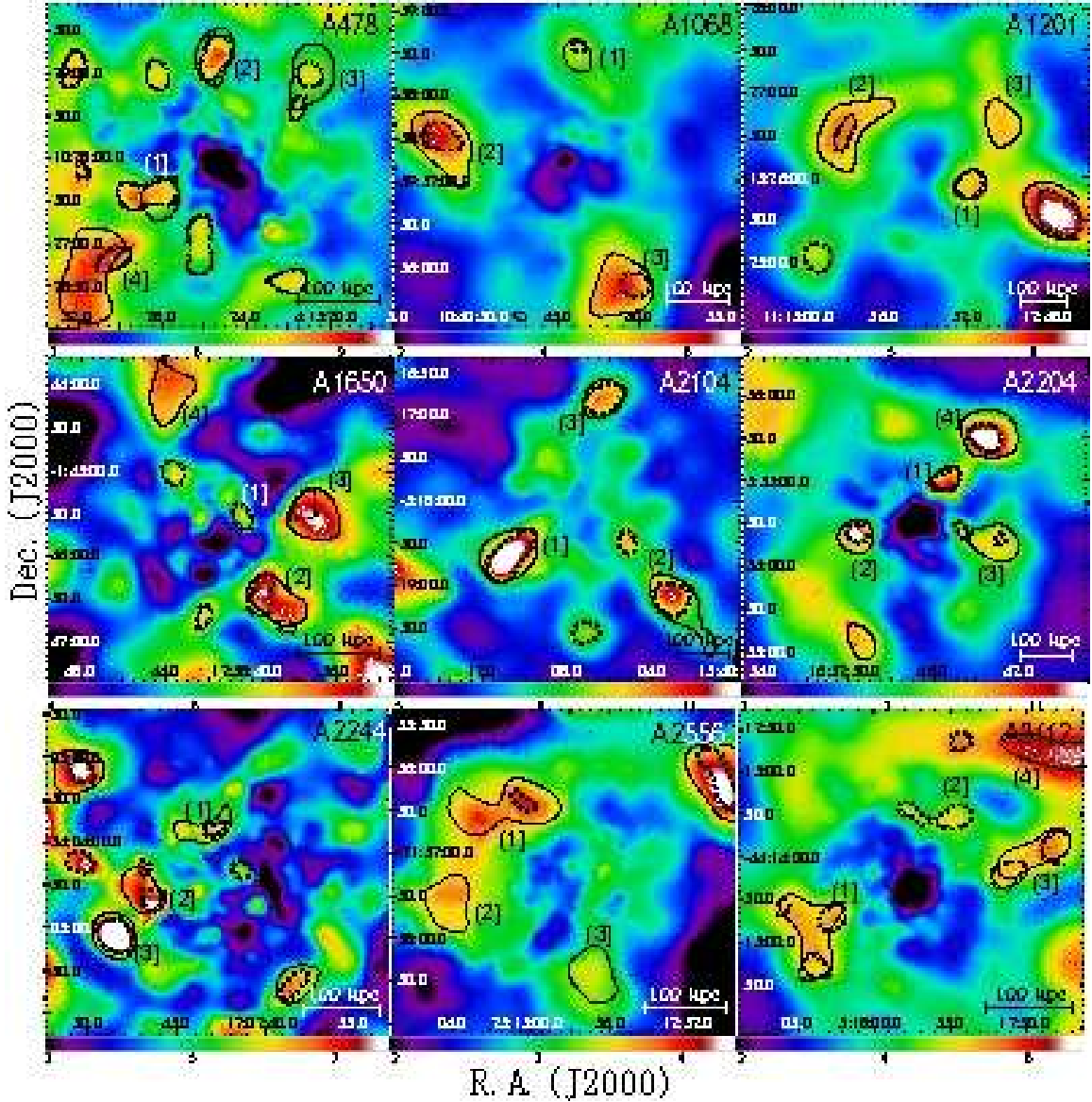}
\caption{Projected temperature maps, on which the most significant
substructures with the characteristic scales indicated by the wavelet spectra (Fig. 5), i.e.,
$i=7$ ($100-200$ $h_{70}^{-1}$
kpc; solid line) and $i=6$ ($50-100$ $h_{70}^{-1}$
kpc; dotted line) are marked (\S 3.3.2). The representative substructures
are labeled with the same numbers as used in Table 4.}
\end{center}
\end{figure}

\begin{figure}
\begin{center}
\includegraphics[angle=0,scale=.60]{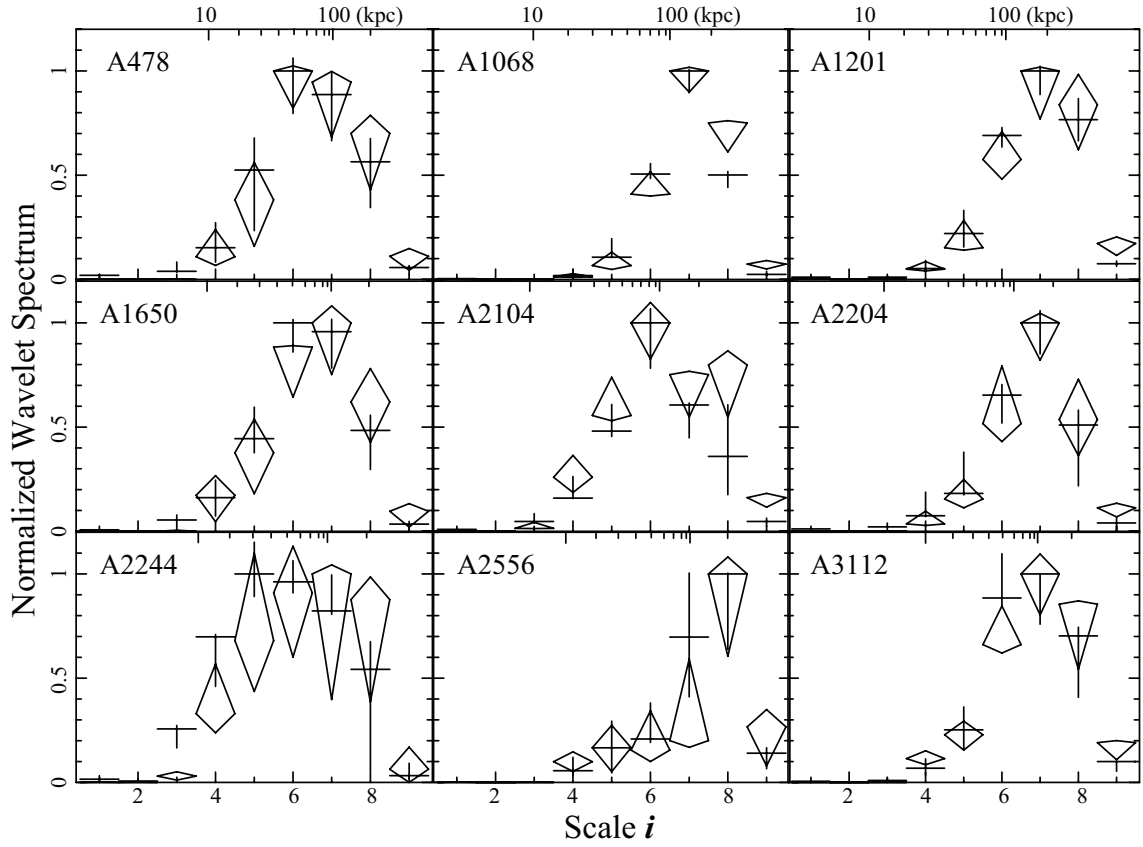}
\caption{Wavelet spectra of the two-dimensional temperature distributions $T({\bf r})$
derived with both the Gaussian wavelet (cross) and the
B-spline wavelet (diamond). Error bars are given at 1$\sigma$
confidence level (\S 3.3.2).}
\end{center}
\end{figure}

\begin{figure}
\epsscale{1.0}
\begin{center}
\includegraphics[angle=0,scale=.60]{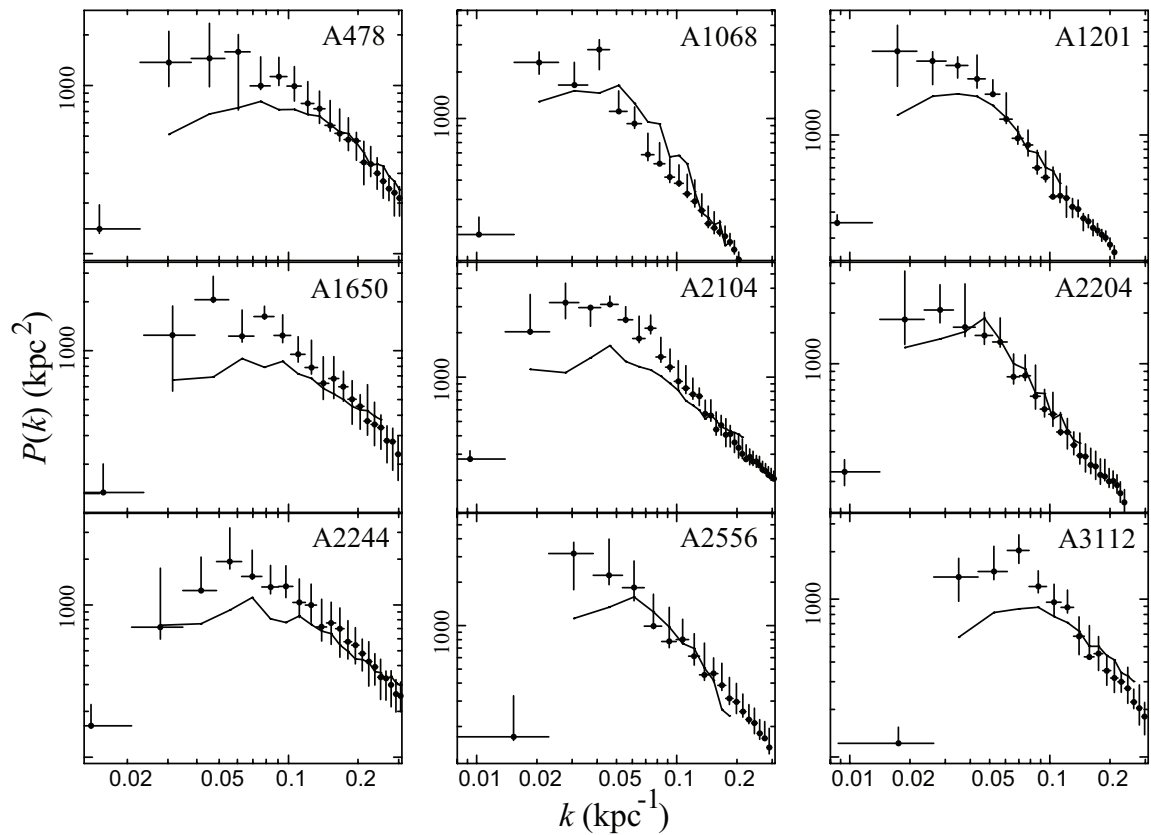}
\caption{
Power spectra of the two-dimensional temperature distributions 
$T({\bf r})$ (cross) and simulated reference temperature maps 
$T_{\rm ref}({\bf r})$ (solid line). $1\sigma$ error bars are determined 
by applying the Monte-Carlo approach introduced in Schuecker et al. 
(2004).}
\end{center}
\end{figure}


\begin{figure}
\epsscale{1.0}
\begin{center}
\includegraphics[angle=0,scale=.60]{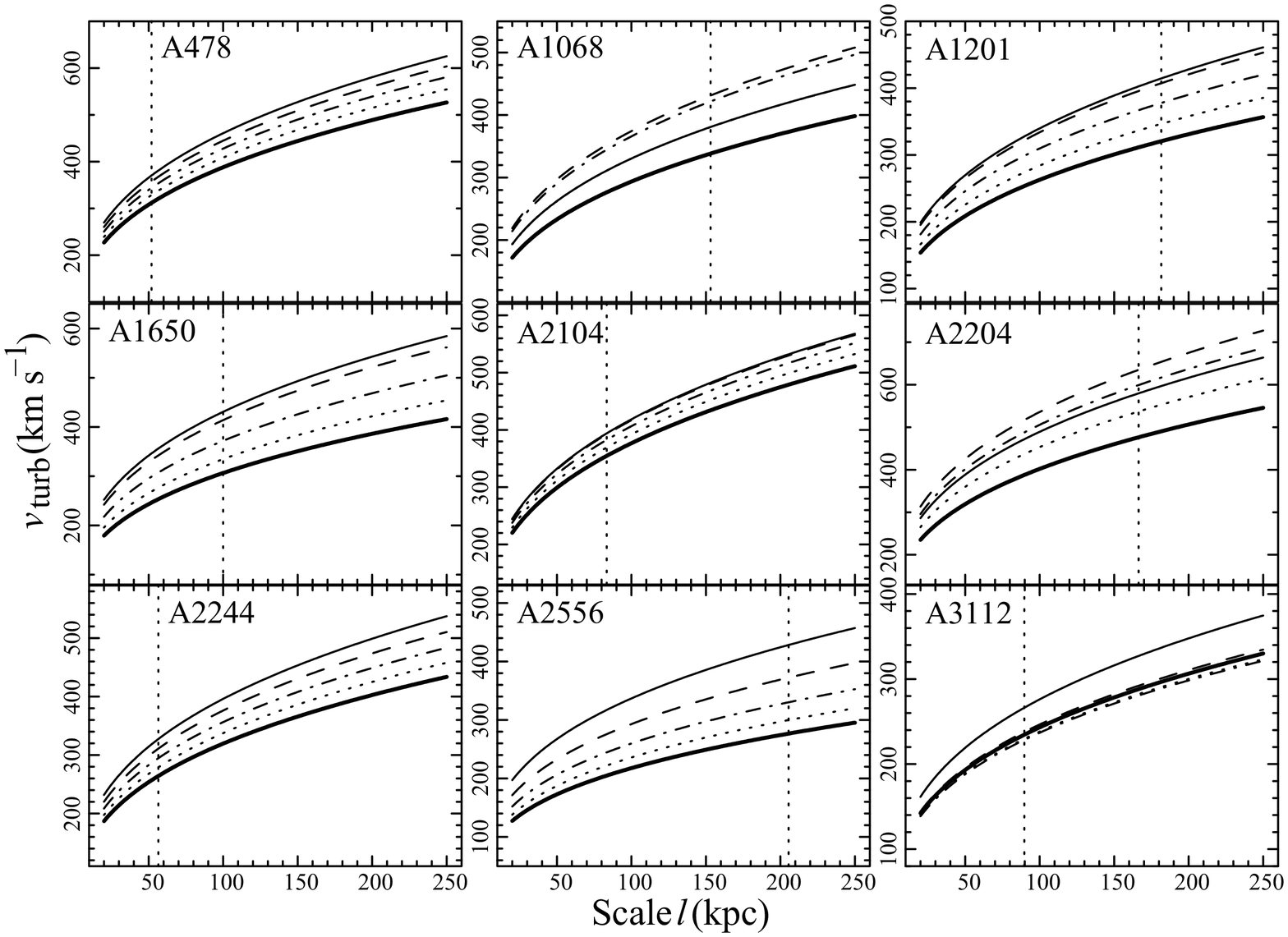}
\caption{
  Turbulence velocities $v_{\rm turb}$ as a function of turbulence length scales $l$, which are
plotted for the radius of 100 (solid), 125 (dash), 150 (dash dot), 175 (dot), and
200 (thick solid) $h_{70}^{-1}$ kpc, respectively.
The vertical lines mark the turnover points obtained in the wavelet spectra (\S 4.1 and Fig. 5).}
\end{center}
\end{figure}



\begin{references}
\reference{} Abell, G. O., Corwin, H. G., \& Olowin, R. P. 1989, ApJS, 70, 1
\reference{} Arnaud, M., Pointecouteau, E., \& Pratt, G. W. 2005, A\&A, 441, 893
\reference{} Bauer, F. E., Fabian, A. C., Sanders, J. S., Allen, S. W., \& Johnstone, R. M. 2005, MNRAS, 359, 1481
\reference{} Biviano, A., Durret, F., Gerbal, D., Le Fevre, O., Lobo, C., Mazure, A., \& Slezak, E. 1996, A\&A, 311, 95
\reference{} Blanton, E. L., Sarazin, C. L., \& McNamara, B. R. 2003, ApJ, 585, 227
\reference{} Bourdin, H., Sauvageot, J.-L., Slezak, E., Bijaoui, A., \& Teyssier, R. 2004, A\&A, 414, 429
\reference{} Bourdin, H., \& Mazzotta, P. 2008, A\&A, 479, 307
\reference{} Bregman, J. N., \& David, L. P. 1989, ApJ, 341, 49
\reference{} Burns, J. O. 1990, AJ, 99, 14
\reference{} Cho, J., Lazarian, A., Honein, A., Knaepen, B., Kassinos, S., \& Moin, P. 2003, ApJ, 589, L77
\reference{} Churazov, E., Br$\ddot{\rm u}$ggen, M., Kaiser, C. R., B$\ddot{\rm o}$hringer, H., \& Forman, W. 2001, ApJ, 554, 261
\reference{} Croston, J. H., Hardcastle, M. J., \& Birkinshaw, M. 2005, MNRAS, 357, 279
\reference{} Dalla Vecchia, C., Bower, R. G., Theuns, T., Balogh, M. L., Mazzotta, P., \& Frenk, C. S. 2004, MNRAS, 355, 995
\reference{} Damiani, F., Maggio, A., Micela, G., \& Sciortino, S. 1997, ApJ, 483, 350
\reference{} Dennis, T. J., \& Chandran, B. D. G. 2005, ApJ, 622, 205
\reference{} Dickey, J. M., \& Lockman, F. J. 1990, ARA$\&$A, 28, 215
\reference{} Donahue, M., Voit, G. M., O'Dea, C. P., Baum, S. A., \& Sparks, W. B. 2005, ApJ, 630, L13
\reference{} Edge, A. C., \& Stewart, G. C. 1991, MNRAS, 252, 414
\reference{} Eilek, J. A. 2003, Phys. Plasmas, 10, 1539
\reference{} Eke, V. R., Cole, S., Frenk, C. S., \& Patrick H., J. 1998, MNRAS, 298, 1145
\reference{} El-Zant, A. A., Kim, W.-T., \& Kamionkowski, M. 2004, MNRAS, 354, 169
\reference{} Evrard, A. E., \& Henry, J. P. 1991, ApJ, 383, 95
\reference{} Evrard, A. E., Metzler, C. A., \& Navarro, J. F. 1996, ApJ, 469, 494
\reference{} Fabian, A. C. 1994a, ARA\&A, 32, 277
\reference{} Fabian, A. C., Crawford, C. S., Edge, A. C., \& Mushotzky, R. F. 1994b, MNRAS, 267, 779
\reference{} Fabian, A. C., Sanders, J. S., Ettori, S., Taylor, G. B., Allen, S. W., Crawford, C. S., Iwasawa, K., Johnstone, R. M., \& Ogle, P. M. 2000, MNRAS, 318, L65
\reference{} Fabian, A. C., Sanders, J. S., Taylor, G. B., \& Allen, S. W. 2005, MNRAS, 360, L20
\reference{} Frenk, C. S., Baugh, C. M., \& Cole, S. M. 1996, IAU Symposia 171, 247
\reference{} Fujita, Y., Matsumoto, T., \& Wada, K. 2004, ApJ, 612, L9
\reference{} Gardini, A. 2007, A\&A, 464, 143
\reference{} Govoni, F., Markevitch, M., Vikhlinin, A., VanSpeybroeck, L., Feretti, L., \& Giovannini, G. 2004, ApJ, 605, 695
\reference{} Grebenev, S. A., Forman, W., Jones, C., \& Murray, S. 1995, ApJ, 445, 607
\reference{} Grevesse, N., \& Sauval, A. J. 1998, Space Sci. Rev., 85, 161
\reference{} Horner, D. J., Mushotzky, R. F., \& Scharf, C. A. 1999, ApJ, 520, 78
\reference{} Ikebe, Y., Ezawa, H., Fukazawa, Y., Hirayama, M., Ishisaki, Y., Kikuchi, K., Kubo, H., Makishima, K., Matsushita, K., Ohashi, T., Takahashi, T., \& Tamura, T. 1996, Nature, 379, 427
\reference{} Jetha, N. N., Ponman, T. J., Hardcastle, M. J., \& Croston, J. H. 2007, MNRAS, 376, 193
\reference{} Kaiser, N. 1991, ApJ, 383, 104
\reference{} Kanov, K. N., Sarazin, C. L., \& Hicks, A. K. 2006, ApJ, 653, 184
\reference{} Kauffmann, G., White, S. D. M., \& Guiderdoni, B. 1993, MNRAS, 264, 201
\reference{} Kim, W.-T., \& Narayan, R. 2003, ApJ, 596, 889
\reference{} Kotov, O., \& Vikhlinin, A. 2006, ApJ, 641, 752
\reference{} Makishima, K., Ezawa, H., Fukuzawa, Y., Honda, H., Ikebe, Y., Kamae, T., Kikuchi, K., Matsushita, K., Nakazawa, K., Ohashi, T., Takahashi, T., Tamura, T., \& Xu, H. 2001, PASJ, 53, 401
\reference{} Markevitch, M., \& Vikhlinin, A. 2001, ApJ, 563, 95
\reference{} Markevitch, M., Mazzotta, P., Vikhlinin, A., Burke, D., Butt, Y., David, L., Donnelly, H., Forman, W. R., Harris, D., Kim, D.-W., Virani, S., \& Vrtilek, J. 2003, ApJ, 586, L19
\reference{} Markevitch, M., \& Vikhlinin, A. 2007, Phys. Rep., 443, 1
\reference{} Maron, J., Chandran, B. D., \& Blackman, E. 2004, Phys. Rev. Lett., 92, 045001
\reference{} Maughan, B. J., Ellis, S. C., Jones, L. R., Mason, K. O., C$\rm \acute{o}$rdova, F. A., \& Priedhorsky, W. 2006, ApJ, 640, 219
\reference{} McNamara, B. R., Wise, M. W., \& Murray, S. S. 2004, ApJ, 601, 173
\reference{} McNamara, B. R., \& Nulsen, P. E. J. 2007, ARA\&A, 45, 117
\reference{} Mohr, J. J., \& Evrard, A. E. 1997, ApJ, 491, 38
\reference{} Mushotzky, R. F. 1984, Phys. Scr., 7, 157
\reference{} Narayan, R., \& Medvedev, M. V. 2001, ApJ, 562, L129
\reference{} Navarro, J. F., Frenk, C. S., \& White, S. D. M. 1995, MNRAS, 275, 720
\reference{} Nulsen, P. E. J., David, L. P., McNamara, B. R., Jones, C., Forman, W. R., \& Wise, M. 2002, ApJ, 568, 163
\reference{} O'Sullivan, E., Vrtilek, J. M., Kempner, J. C., David, L. P., \& Houck, J. C. 2005, MNRAS, 357, 1134
\reference{} Peterson, J. R., Kahn, S. M., Paerels, F. B. S., Kaastra, J. S., Tamura, T., Bleeker, J. A. M., Ferrigno, C., \& Jernigan, J. G. 2003, ApJ, 590, 207
\reference{} Petrosian, V., Bykov, A., \& Rephaeli, Y. 2008, Space Sci. Rev., 134, 191
\reference{} Pizzolato, F., \& Soker, N. 2006, MNRAS, 371, 1835
\reference{} Pointecouteau, E., Arnaud, M., Kaastra, J., \& de Plaa, J. 2004, A$\&$A, 423, 33
\reference{} Ponman, T. J., Cannon, D. B., \& Navarro, J. F. 1999, Nature, 397, 135
\reference{} Ponman, T. J., Sanderson, A. J. R., \& Finoguenov, A. 2003, MNRAS, 343, 331
\reference{} Poole, G. B., Fardal, M. A., Babul, A., McCarthy, I. G., Quinn, T., \& Wadsley, J. 2006, MNRAS, 373, 881
\reference{} Quilis, V., Bower, R. G., \& Balogh, M. L. 2001, MNRAS, 328, 1091
\reference{} Rebusco, P., Churazov, E., B$\ddot{\rm o}$hringer, H., \& Forman, W. 2006, MNRAS, 372, 1840
\reference{} Rood, H. J., \& Sastry, G. N. 1971, PASP, 83, 313
\reference{} Rosati, P., della Ceca, R., Burg, R., Norman, C., \& Giacconi, R. 1995, ApJ, 445, L11
\reference{} Sanders, J. S., Fabian, A. C., Allen, S. W., \& Schmidt, R. W. 2004, MNRAS, 349, 952  
\reference{} Sanders, J. S., Fabian, A. C., \& Taylor, G. B. 2005, MNRAS, 356, 1022
\reference{} Sanders, J. S., Fabian, A. C., \& Taylor, G. B. 2009, MNRAS, 393, 71
\reference{} Sanderson, A. J. R., Finoguenov, A., \& Mohr, J. J. 2005, ApJ, 630, 191
\reference{} Sanderson, A. J. R., Ponman, T. J., \& O'Sullivan, E. 2006, MNRAS, 372, 1496
\reference{} Sarazin, C. L. 1999, ApJ, 520, 529
\reference{} Schombert, J. M., West, M. J., Zucker, J. R., \& Struble, M. F. 1989, AJ, 98, 1999
\reference{} Schuecker, P., Finoguenov, A., Miniati, F., B$\ddot{\rm o}$hringer, H., \& Briel, U. G. 2004, A\&A, 426, 387
\reference{} Sharon, K., Gal-Yam, A., Maoz, D., Filippenko, A. V., \& Guhathakurta, P. 2007, ApJ, 60, 1165
\reference{} Shensa, M. J. 1992, IEEE Trans. Sig. Proc., 40, 2464
\reference{} Shibata, R., Matsushita, K., Yamasaki, N. Y., Ohashi, T., Ishida, M., Kikuchi, K., B$\ddot{\rm o}$hringer, H., \& Matsumoto, H. 2001, ApJ, 549, 228
\reference{} Slezak, E., Durret, F., \& Gerbal, D. 1994, AJ, 108, 1996
\reference{} Snowden, S. L., Egger, R., Finkbeiner, D., Freyberg, M. J., \& Plucinsky, P. P. 1998, ApJ, 493, 715
\reference{} Spitzer, L. 1962, Physics of Fully Ionized Gases (New York: Interscience)
\reference{} Spitzer, L. 1978, Physical Processes in the Interstellar Medium (New York: John Wiley \& Sons)
\reference{} Starck, J.-L., Murtagh, F., \& Bijaoui, A. 1995, ADASS Conf. Proc. ASP Conf. Series, 77, 279
\reference{} Starck, J.-L., \& Pierre, M. 1998, A$\&$AS, 128, 397
\reference{} Starck, J. L., Donoho, D. L., \& Cand$\grave{\rm e}$s, E. J. 2003, A\&A, 398, 785
\reference{} Struble, M. F., \& Rood, H. J. 1987, ApJS, 63, 555
\reference{} Sun, M., Jones, C., Murray, S. S., Allen, S. W., Fabian, A. C., \& Edge, A. C. 2003, ApJ, 587, 619
\reference{} Takahashi, S., \& Yamashita, K. 2003, PASJ, 55, 1105
\reference{} Takizawa, M., Sarazin, C. L., Blanton, E. L., \& Taylor, G. B. 2003, ApJ, 595, 142
\reference{} Thornton, K., Gaudlitz, M., Janka, H.-Th., \& Steinmetz, M. 1998, ApJ, 500, 95
\reference{} Torrence, C., \& Compo, G. P. 1998, Bull. Amer. Met. Soc., 79, 61
\reference{} Tozzi, P., \&  Norman, C. 2001, ApJ, 546, 63
\reference{} Van Cittert, P. H. 1931, Z. Phys., 69, 298
\reference{} Vernaleo, J. C., \& Reynolds, C. S., 2006, ApJ, 645, 83
\reference{} Vikhlinin, A., Forman, W., \& Jones, C. 1997, ApJ, 474, L7
\reference{} Vikhlinin, A., McNamara, B. R., Forman, W., Jones, C., Quintana, H., \& Hornstrup, A. 1998, ApJ, 502, 558
\reference{} Vikhlinin, A., Markevitch, M., \& Murray, S. S. 2001, ApJ, 551, 160
\reference{} Vikhlinin, A., Markevitch, M., Murray, S. S., Jones, C., Forman, W., \& Van Speybroeck, L. 2005, ApJ, 628, 665
\reference{} Vikhlinin, A., Kravtsov, A., Forman, W., Jones, C., Markevitch, M., Murray, S. S., \& Van Speybroeck, L. 2006, ApJ, 640, 691
\reference{} Voigt, L. M.. \& Fabian, A. C. 2004, MNRAS, 347, 1130
\reference{} Wise, M. W., McNamara, B. R., \& Murray, S. S. 2004, ApJ, 601, 184
\reference{} Xu, H., Jin, G., \& Wu, X.-P. 2001, ApJ, 553, 78
\reference{} Zakamska, N. L., \& Narayan, R. 2003, ApJ, 582, 162


\reference{}
\reference{}
\reference{}
\reference{}
\reference{}
\reference{}
\reference{}
\reference{}
\reference{}
\reference{}
\reference{}
\reference{}
\reference{}
\reference{}
\reference{}
\reference{}
\reference{}
\reference{}
\reference{}
\reference{}
\reference{}
\reference{}
\reference{}
\reference{}
\reference{}
\reference{}
\reference{}
\reference{}
\reference{}
\reference{}
\reference{}
\reference{}
\reference{}
\reference{}
\reference{}
\reference{}
\reference{}
\reference{}
\reference{}
\reference{}
\reference{}
\reference{}
\reference{}
\reference{}
\reference{}
\reference{}
\reference{}
\reference{}
\reference{}
\reference{}
\reference{}
\reference{}
\reference{}
\reference{}
\reference{}
\reference{}
\reference{}
\reference{}
\reference{}
\reference{}
\reference{}
\reference{}
\reference{}
\reference{}
\reference{}
\reference{}
\reference{}
\reference{}
\reference{}
\reference{}
\reference{}
\reference{}
\reference{}
\reference{}
\reference{}
\reference{}
\reference{}
\reference{}
\reference{}
\reference{}
\reference{}
\reference{}
\reference{}
\reference{}
\reference{}
\reference{}
\reference{}
\reference{}
\reference{}
\reference{}
\reference{}
\reference{}
\reference{}

\end{references}
\end{document}